\documentclass{article}

\usepackage{microtype}
\usepackage{graphicx}
\usepackage{subcaption}
\usepackage{booktabs} 
\usepackage{multirow} 
\usepackage[table,dvipsnames]{xcolor} 
\usepackage{bm}
\usepackage{xspace}
\usepackage{hyperref}

\usepackage[preprint]{icml2026}

\usepackage[T1]{fontenc}
\usepackage[utf8]{inputenc}
\usepackage{amsmath}
\usepackage{amssymb}
\usepackage{enumitem}
\usepackage{mathtools}
\usepackage{amsthm}
\usepackage{algorithm}

\usepackage{algpseudocode}

\usepackage[capitalize,noabbrev]{cleveref}

\theoremstyle{plain}

\theoremstyle{definition}

\theoremstyle{remark}

\usepackage[disable,textsize=tiny]{todonotes}

\icmltitlerunning{Not All Prefills Are Equal: PPD Disaggregation for Multi-turn LLM Serving}

\begin{document}

\twocolumn[
\icmltitle{Not All Prefills Are Equal: \\ PPD Disaggregation for Multi-turn LLM Serving}

\icmlsetsymbol{equal}{*}

\begin{icmlauthorlist}
\icmlauthor{Zongze Li}{uchi}
\icmlauthor{Jingyu Liu}{uchi}
\icmlauthor{Zhen Xu}{uchi}
\icmlauthor{Yineng Zhang}{ind}
\icmlauthor{Tahseen Rabbani}{uchi}
\icmlauthor{Ce Zhang}{uchi}
\end{icmlauthorlist}

\icmlaffiliation{uchi}{University of Chicago, USA}
\icmlaffiliation{ind}{Independent Researcher}

\icmlcorrespondingauthor{Zongze Li}{zongzel@uchicago.edu}

\icmlkeywords{Large Language Models, Inference Optimization, Disaggregated Serving, Multi-turn Conversations}

\vskip 0.3in
]

\printAffiliationsAndNotice{}

\begin{abstract}
Prefill-Decode (PD) disaggregation has become the standard architecture for modern LLM inference engines, which alleviates the interference of two distinctive workloads.
With the growing demand for multi-turn interactions in chatbots and agentic systems, we re-examined PD in this case and found two fundamental inefficiencies: (1) every turn requires prefilling the new prompt and response from the last turn, and (2) repeated KV transfers between prefill and decode nodes saturate the bandwidth, leading to high latency and even service degradation.
Our key insight is that not all prefill operations are equally disruptive: \textit{append-prefill}, which processes only the new input tokens while reusing cached KV states, incurs an order-of-magnitude smaller decoding slowdown than full prefill.
This motivates routing append-prefill to decode nodes locally. However, through comprehensive analysis, we show that \textit{no single fixed routing strategy satisfies all Service Level Objectives (SLOs) simultaneously}.
Based on this insight, we propose \textbf{P}refill \textbf{P}refill-capable \textbf{D}ecode (PPD) disaggregation, a dynamic routing system that decides when to process Turn 2+ requests locally on decode nodes using cached KV states.
PPD adapts to varying SLOs via configurable weights and seamlessly integrates with traditional PD deployments.
With extensive evaluations, we show that PPD reduces Turn 2+ time-to-first-token (TTFT) by $\sim$68\% while maintaining competitive time-per-output-token (TPOT), effectively alleviating KV transfer congestion under high load.
PPD provides a flexible and efficient paradigm for multi-turn LLM serving.
\end{abstract}

\section{Introduction}
\label{sec:intro}

Prefill-decode (PD) disaggregation has become the standard architecture for LLM inference~\cite{zhong2024distservedisaggregatingprefilldecoding, patel2024splitwiseefficientgenerativellm, deepseekai2025deepseekv3technicalreport, sun2024llumnixdynamicschedulinglarge}, which places compute-intensive prompt processing (prefill) and memory-bound token generation (decode) separately on dedicated GPU pools. 
Despite being effective for independent single-turn queries, PD exhibits critical inefficiencies under \textit{multi-turn conversations}, the dominant usage pattern in chatbots and agentic systems~\cite{duan2023botchatevaluatingllmscapabilities}. 
When a typical PD handles a new-turn query, PD routes the prompt to the prefill node along with the response string from the previous turn. A subtle but consequential property of canonical PD designs is that KV transfer is strictly P$\to$D: P operates as a producer, D as a consumer, with no reverse path. Even though the response token KV from the previous turn already resides on D, it is inaccessible to P; the prefill node must therefore recompute the KV cache for the entire conversation history (prior responses plus the new prompt) before transferring it back to D. Recent measurements on real chat workloads find that this recomputation accounts for up to 99\% of multi-turn prefill cost~\cite{gao2024costefficientlargelanguagemodel}.
We re-examine PD under the case where multi-turn conversations are prevalent, and find that the standard PD strategy causes time-to-first-token (TTFT) for Turn 2+ (the second turn and beyond) to remain high even with shorter inputs, and that these repeated KV transfers saturate network bandwidth (\Cref{sec:validation}).

Our key insight is that the behavior of prefill operations differs sharply between the first and later turns, motivating a specialized strategy. In \Cref{fig:interference-tpot}, we demonstrate that \textit{not all prefill operations are equally disruptive for decoding}: full prefill (a new prompt without cached context) causes an order of magnitude greater decode slowdown than \textit{append-prefill} (only the new input tokens, reusing cached KV).
This order-of-magnitude gap indicates that running append-prefill (AP) on decode nodes locally can efficiently handle Turn 2+ prompts with minimal interference.
While a concurrent work~\cite{he2026efficientmultiroundllminference} shares the high-level intuition of routing incremental prefills to decode nodes, our work uniquely grounds this approach in a rigorous micro-architectural interference analysis and formalizes the routing decision as an optimization problem.

On the one hand, routing AP operations to prefill nodes can maintain a high TPOT; on the other hand, routing to the decode nodes can improve overall throughput. The central question is therefore \textit{should we route the AP operations to the prefill or decode nodes}? We formalize this problem with a common framework in which PD is a special case that always routes AP to the prefill nodes. To understand the trade-offs (see \Cref{sec:tradeoff}), we routed a different fixed percentage of AP to decode nodes for each strategy, and we found \textit{no single fixed strategy meets all Service Level Objectives (SLOs)} (see \Cref{tab:winner-distribution}).

To this end, we introduce \textbf{P}refill \textbf{P}refill-capable \textbf{D}ecode (PPD) disaggregation (\Cref{sec:ppd-design}), which proposes to dynamically route AP operations to the decode nodes based on the current workload estimate, user-specified SLO, and initial node assignment. PPD optimizes the disaggregated serving objective with a simple yet effective algorithm using precomputed offline statistics, and can always recover the default PD configuration when it needs to. In the extreme case, PPD can route all AP operations to the decode nodes, where the locally cached KV states can be fully reused without extra recomputation or KV transfers.

\begin{figure*}[t]
    \centering
    \includegraphics[width=0.95\textwidth]{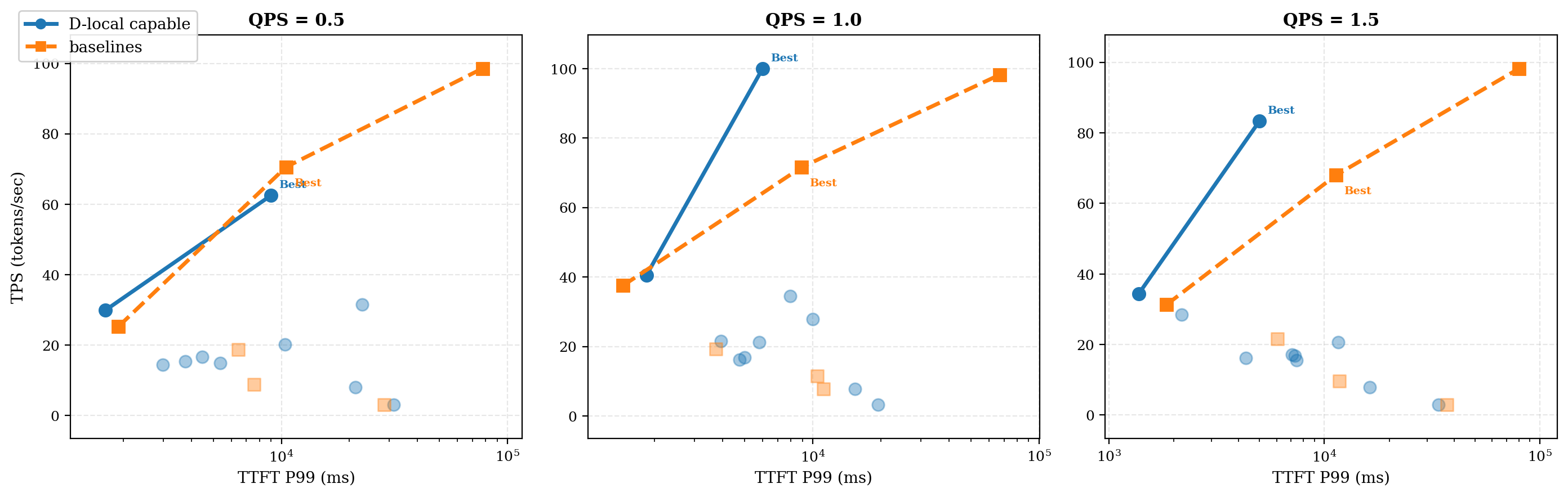}
    \caption{
      \textbf{99th-percentile (P99) TTFT vs.\ Tokens-per-Second (TPS) Pareto frontiers} under a long-context workload (10k input, 100 output tokens, 5 turns) at three load levels. Higher TPS and lower TTFT are better (upper-left is ideal). \textit{Baselines} (orange): PD and Replica configurations where Turn 2+ requests always require P-node processing. \textit{D-local capable} (blue): configurations allowing decode nodes to process Turn 2+ locally via append-prefill. The ``Best'' annotation marks the configuration selected by our dynamic PPD routing system, validating correct trade-off optimization. See \Cref{fig:pareto-small-context} in the Appendix for small-context results showing consistent trends across different workloads.
} 
    \label{fig:pareto-overview}
\end{figure*}

On both synthetic and real datasets, we evaluate the effectiveness of our proposed PPD framework and show that PPD achieves the best Pareto frontier compared to fixed routing or the default PD (see \Cref{fig:pareto-overview}). We also conduct detailed ablation studies to justify our design choices and analyze complex serving scenarios with varying requirements. We summarize our main contributions as follows.
\begin{enumerate}[label=\textbf{(\roman*)},wide=0pt,labelsep=0.5em,itemsep=6pt,topsep=2pt,parsep=0pt]
    \item We identify the inefficiencies of PD in multi-turn conversations (\Cref{sec:interference}): running AP on the prefill node incurs extra recomputation and saturates network bandwidth via frequent KV cache transfers.
    
    \item We find that full prefill and AP differ by an order of magnitude in decode interference (48\% vs.\ 2\% TPOT slowdown at batch size 200), motivating selective AP-to-D routing.
    
    \item We formalize multi-turn inference serving as an optimization problem in which traditional PD is the special case $x{\equiv}0$, and show that no single fixed strategy dominates (\Cref{subsec:no-universal-best}).
    
    \item We propose PPD, a dynamic routing system that selects $x$ per request from current workload, operator weights, and initial node assignment.
    
    \item Detailed evaluations (\Cref{sec:validation}) show PPD outperforms standard PD and fixed strategies, achieving 48--73\% Turn 2+ TTFT reduction on synthetic sweeps while maintaining competitive TPOT.
\end{enumerate}

\section{Background}
\label{sec:background}

\subsection{LLM Inference and Scaling}

LLM inference proceeds in two phases.
\textit{Prefill} processes the input prompt in parallel,
producing the first output token and populating the KV
cache; it is compute-bound with $O(n^2)$ attention complexity for
$n$ tokens~\cite{liu2025speculativeprefillturbochargingttft,shi2024discoveringgemsearlylayers}, though FlashAttention~\cite{dao2022flashattentionfastmemoryefficientexact, shah2024flashattention3fastaccurateattention} reduces memory complexity from quadratic to linear.
\textit{Decode} then generates tokens autoregressively, loading
the full model weights for each token; it is
memory-bandwidth-bound~\cite{280922, kwon2023efficientmemorymanagementlarge}. Grouped Query Attention (GQA)~\cite{ainslie2023gqatraininggeneralizedmultiquery} reduces the KV cache memory footprint while maintaining model quality.
The KV cache (key-value states from prior tokens) grows linearly with context and, alongside model weights, dominates GPU memory.

The most straightforward approach to scaling LLM serving is
\textit{replication}: deploying identical model instances across
GPUs, each handling requests independently.
However, co-locating prefill and decode on the same GPU leads to
\textit{prefill-decode interference}~\cite{zhong2024distservedisaggregatingprefilldecoding}: long-running prefill
computations block decode iterations, causing unpredictable latency
spikes for ongoing generation.
Chunked-prefill techniques~\cite{298679} mitigate this
by splitting prefill into smaller chunks that interleave with decode;
DeepSpeed-FastGen~\cite{holmes2024deepspeedfastgenhighthroughputtextgeneration} further introduces Dynamic SplitFuse for efficient token composition.
However, these techniques cannot fully eliminate interference when prefill workloads are
heavy (we quantify this gap in \Cref{sec:interference}).

\subsection{Prefill-Decode Disaggregation}
Prefill-decode (PD) disaggregation mitigates interference by physically separating prefill and decode onto distinct GPU pools~\cite{zhong2024distservedisaggregatingprefilldecoding, patel2024splitwiseefficientgenerativellm}.
Prefill nodes (P) process incoming prompts and transfer the
resulting KV cache to decode nodes (D) over the network; D then
performs autoregressive generation without interruption.
This architecture eliminates interference, enables independent
scaling of P and D resources, and permits hardware
heterogeneity~\cite{patel2024splitwiseefficientgenerativellm}.

PD disaggregation has rapidly become the industry standard.
It is supported by all major serving frameworks (vLLM~\cite{kwon2023efficientmemorymanagementlarge},
SGLang~\cite{zheng2024sglangefficientexecutionstructured}, TensorRT-LLM, LMDeploy~\cite{2023lmdeploy}, and NVIDIA
Dynamo~\cite{nvidia2025dynamo}) and is deployed at production
scale by providers such as DeepSeek~\cite{deepseekai2025deepseekv3technicalreport} and Gemini~\cite{geminiteam2025geminifamilyhighlycapable}.
However, disaggregation introduces a fundamental cost: every
request requires transferring the full KV cache over the network.
For Llama-3.1-8B with a 2K-token context, each transfer is
$\sim$256~MB.
Recent work explores alternative disaggregation strategies: DuetServe~\cite{gao2025duetserveharmonizingprefilldecode} achieves disaggregation-level isolation within a single GPU through adaptive SM partitioning; Nexus~\cite{shi2025nexusproactiveintragpudisaggregationprefill} proactively reallocates GPU resources between phases; and TaiChi~\cite{wang2025prefilldecodeaggregationdisaggregationunifying} unifies aggregation and disaggregation, showing that the optimal strategy depends on SLO constraints.

\paragraph{One-way KV Transfer Protocol.}
A defining architectural property of PD disaggregation is the directionality of its KV transfer protocol: P nodes act as KV \emph{producers} and D nodes as \emph{consumers}, with no reverse channel from D back to P. This producer/consumer contract is preserved across all major production engines~\cite{zhong2024distservedisaggregatingprefilldecoding, kwon2023efficientmemorymanagementlarge, zheng2024sglangefficientexecutionstructured}. A direct consequence in multi-turn serving is that any KV state computed on D, including all decoded responses, is unreachable from P, so each new turn re-traverses the full P$\to$D pipeline (\Cref{sec:intro}). External KV cache layers~\cite{qin2025mooncakekvcachecentricdisaggregatedarchitecture, hu2024memservecontextcachingdisaggregated, gao2024costefficientlargelanguagemodel} address this by adding a \emph{separate} storage tier outside the disaggregation topology, rather than altering the producer/consumer contract itself.

\subsection{Multi-turn Serving and KV Cache Reuse}

Real-world LLM deployments are dominated by multi-turn
conversations: chatbots and agentic systems typically involve
multiple turns per session~\cite{10.1145/3774909}.
Under PD disaggregation, each new turn is sent to a P node, which re-computes the entire conversation's KV cache (including prior outputs), then transfers it to a decode node for autoregressive generation.
Historical tokens dominate input length in later turns~\cite{gao2024costefficientlargelanguagemodel}, making this recomputation the dominant prefill cost in multi-turn workloads.

\paragraph{Existing Approaches.}
A growing line of work addresses this via \textit{external KV cache stores}:
CachedAttention~\cite{gao2024costefficientlargelanguagemodel} uses hierarchical caching;
Mooncake~\cite{qin2025mooncakekvcachecentricdisaggregatedarchitecture} provides cluster-wide distributed KV stores;
MemServe~\cite{hu2024memservecontextcachingdisaggregated} unifies context caching with disaggregation;
and LMCache~\cite{liu2025lmcacheefficientkvcache} offers a modular KV cache layer.
Prefix caching in SGLang~\cite{zheng2024sglangefficientexecutionstructured} and vLLM~\cite{kwon2023efficientmemorymanagementlarge} enables KV reuse for repeated prefixes.
Complementary approaches include selective recomputation~\cite{yao2025cacheblendfastlargelanguage}, CPU offloading~\cite{chen2024magicpiglshsamplingefficient, sun2025shadowkvkvcacheshadows}, KV compression~\cite{li2024snapkvllmknowslooking, liu2025hamburgeracceleratingllminference}, streaming~\cite{liu2024cachegenkvcachecompression}, prefix-aware attention~\cite{ye2024chunkattentionefficientselfattentionprefixaware}, and compressive memory~\cite{munkhdalai2024leavecontextbehindefficient}. 

These approaches share a common strategy: storing KV caches externally or requiring requests to return to the same instance.
We take a complementary approach: adjusting \textit{routing} so follow-up turns execute directly on the decode GPU that already holds the KV cache.
A concurrent work, AMPD~\cite{he2026efficientmultiroundllminference}, also explores a routing-based approach to mitigate multi-turn inefficiencies, albeit using real-time queue states rather than our offline optimization framework.
We discuss the relationship with distributed KV stores and concurrent routing systems in \Cref{sec:discuss}.

\section{Dynamic Routing of Append-Prefill Operations as an Optimization Problem}
\label{sec:formulation}

The central question is \emph{where} to compute \textbf{A}ppend-\textbf{P}refill (AP) operations, on prefill or decode nodes, in disaggregated serving. We use $x$ in two related senses: a hardware-level fraction $x \in [0, 1]$ of AP routed to D (uniform across requests; \Cref{sec:tradeoff}), and a per-request binary decision $x \in \{0, 1\}$ (route through P or process locally on D; \Cref{sec:ppd-design}). \Cref{sec:tradeoff} sweeps five static fractions $x \in \{0, \tfrac{1}{3}, \tfrac{1}{2}, \tfrac{2}{3}, 1\}$ and three P:D node assignments $\pi$ (1P\_3D, 2P\_2D, 3P\_1D), under workloads $\psi$ that span decode-heavy, balanced, and prefill-heavy profiles. We focus on pure P/D configurations; hybrid R+P/D configurations generally underperform (\Cref{subsec:config-space}). No single static $x$ dominates TTFT, TPOT, and throughput simultaneously, which motivates a per-request policy.

Given $\pi$ and operator-specified weights $\mathbf{w} = (w_{\text{ttft}}, w_{\text{tpot}})$, we score the benefit of local processing ($x{=}1$) over the PD path ($x{=}0$) for workload $\psi$ as

\vspace{-5pt}
\begin{equation}
    \label{eq:score}
    S(\psi;\, \pi, \mathbf{w}) \;=\; w_{\text{ttft}}\,\Delta_{\text{ttft}} \;-\; w_{\text{tpot}}\,\Delta_{\text{tpot}},
\end{equation}
\vspace{-5pt}

\noindent where $\Delta_{\text{ttft}}$ is the relative TTFT improvement and $\Delta_{\text{tpot}}$ the relative TPOT degradation when $\psi$ is processed locally rather than through P. Each Turn 2+ request is routed locally when $S > 0$ and through P otherwise; Turn 1 always uses the PD path since no cached KV exists. PPD instantiates this policy via an offline-computed lookup table (\Cref{sec:ppd-design}); traditional PD is the special case $x \equiv 0$. Throughput is a system-level emergent metric and is not optimized per-request, but it improves as KV transfer volume drops (\Cref{tab:winner-distribution}).

\section{Interference Analysis and Comprehensive Trade-Offs}
\label{sec:tradeoff}

In this section, we begin by establishing a key empirical result that underpins the benefit of routing AP to D:
Append-prefill causes an order-of-magnitude smaller interference than full prefill, making it feasible for decode nodes to handle Turn 2+ locally.
We then conduct a comprehensive benchmark spanning 3,060
configurations to evaluate the effectiveness of Full AP-to-D configurations ($x{=}1$) and characterize
their trade-offs.
Throughout this section, we analyze static deployments where decode nodes with $x{=}1$ always handle Turn 2+ locally; \Cref{sec:ppd-design} extends this to PPD, our dynamic per-request routing system.

\subsection{Quantifying Prefill Interference}
\label{sec:interference}

The core assumption behind PD disaggregation is that \textit{all}
prefill operations interfere severely with decode, necessitating
physical isolation.
Recent work~\cite{gao2025duetserveharmonizingprefilldecode, shi2025nexusproactiveintragpudisaggregationprefill} has begun to challenge this assumption by exploring adaptive isolation strategies.
We further this direction by distinguishing two types of prefill.

\paragraph{Full Prefill vs.\ Append Prefill.}
\textit{Full prefill} processes a new prompt without any cached context, the standard Turn 1 scenario with $O(n^2)$ attention complexity for $n$ input tokens.
\textit{Append prefill} processes only new tokens while reusing cached KV from previous turns.
For Turn 2+ with $m$ new tokens appended to the context of $n$ cached tokens, append-prefill computes attention only for the $m$ new tokens (each attending over $n+m$ keys), yielding $O(m(n+m))$ complexity.
When $m \ll n$ (typical for follow-ups), append-prefill is roughly $n/m$ times cheaper than full prefill of the same total sequence length.

\begin{figure}[htbp]
    \centering
    \includegraphics[width=0.8\columnwidth]{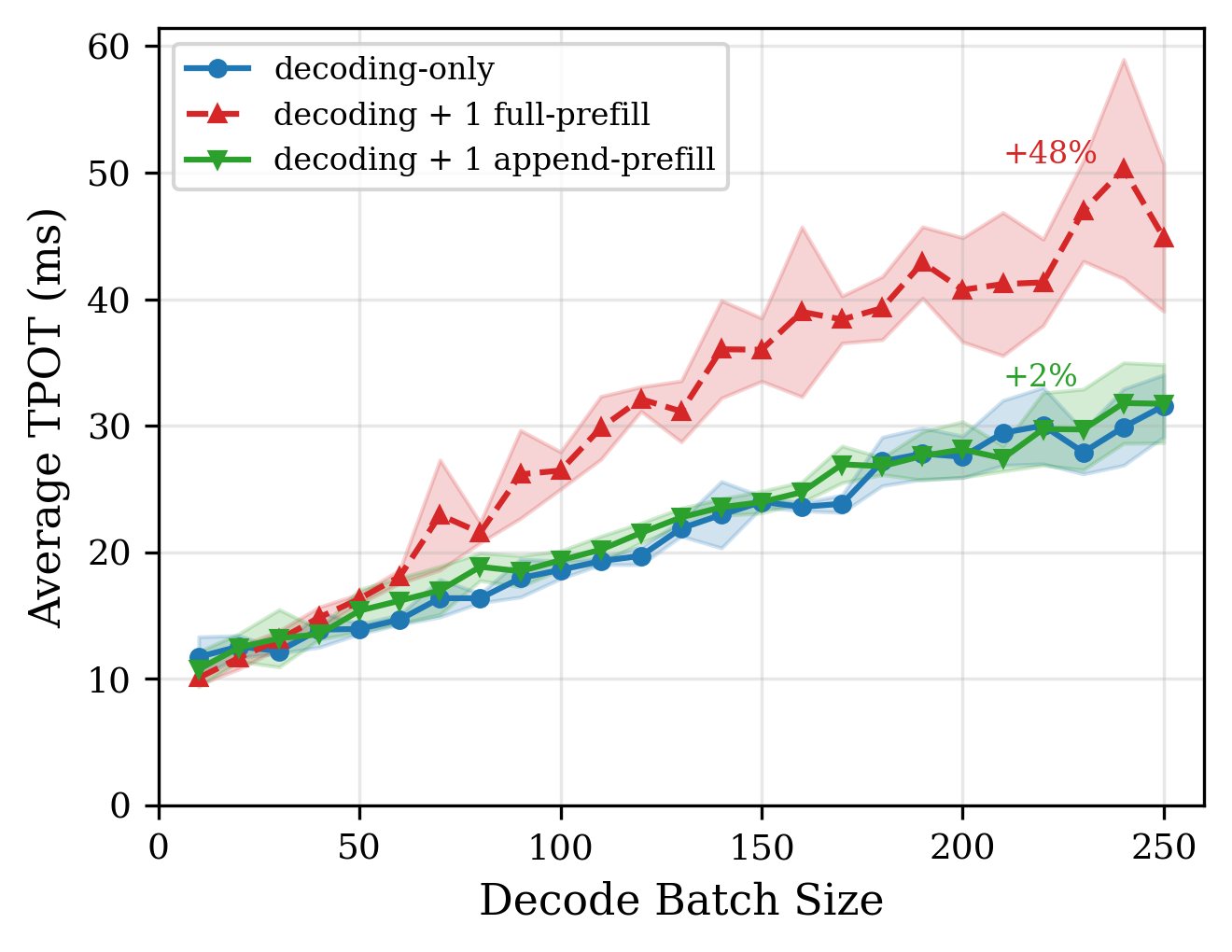}
    \caption{
        \textbf{Prefill-decode interference.}
        Decode TPOT degradation when co-locating with one full-prefill vs.\ one append-prefill operation (both processing 1,024 tokens).
        Full prefill causes a significant slowdown; append-prefill remains close to baseline.
        See \Cref{fig:interference-4prefills} in the Appendix for 4-prefill experiments showing consistent trends.
    }
    \label{fig:interference-tpot}
\end{figure}

\paragraph{Interference Measurement.}
We conduct controlled micro-benchmarks on a single H100 GPU using Llama-3.1-8B to isolate interference effects, following established roofline analysis methodology~\cite{yuan2024llminferenceunveiledsurvey}.
We measure decode TPOT degradation when co-locating with full or append-prefill (\Cref{fig:interference-tpot}).
Full prefill causes $\sim$48\% slowdown at batch size 200; append-prefill causes only $\sim$2\%, an order of magnitude less.
With 4 concurrent prefills (see \Cref{fig:interference-4prefills}), full reaches +57\% while append stays at +21\%, so the gap persists under increased concurrency.

This gap persists across context lengths: full prefill interference grows to 3--4$\times$ at 32K tokens while append-prefill stays below 25\% even at 64K (\Cref{fig:interference-sensitivity} in Appendix).
These results confirm that decode nodes can safely handle Turn 2+ locally.

\subsection{Configuration Space and Methodology}
\label{subsec:config-space}
In this subsection, we describe the configuration space and methodology for systematically evaluating how different P/D assignments and routing parameters affect multi-turn serving performance.
\begin{figure*}[t]
    \centering
    \includegraphics[width=\linewidth]{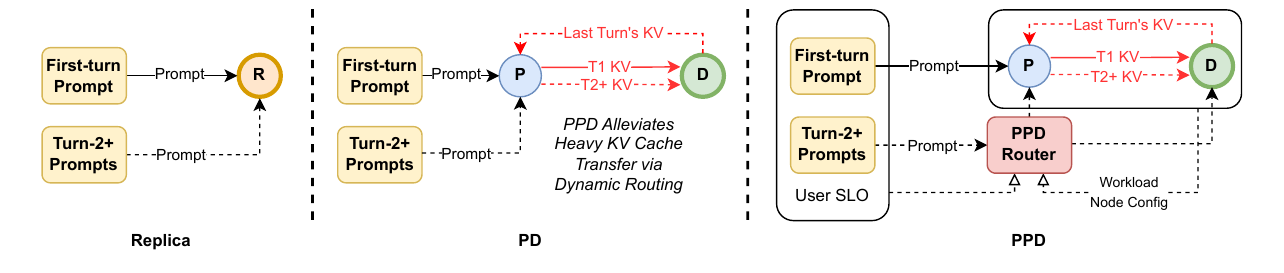}
    \vspace{-20pt}
    \caption{\textbf{Dynamic routing of append-prefill with PPD.} We illustrate the core concept behind PPD: \textbf{PPD} (Right) dynamically routes the append-prefill operations for multi-turn conversations to the prefill or decode nodes based on user SLO, estimated workload from the system, and the initial node configuration. Compared to \textbf{Replica} (Left), PPD retains all the benefits of disaggregation. In contrast to \textbf{PD} (Middle), PPD alleviates heavy KV cache transfer as well as extra recomputation for append-prefill with local cache, and can always adjust to meet various serving requirements. We want to highlight that both PD and any strategies with a fixed $x\%$ of append-prefill routed to decode nodes are special cases of our PPD. }
    \label{fig:configuration}
\end{figure*}

\paragraph{Machine Types and Routing Parameter.}
We define three machine types: \textbf{P} (prefill-only), \textbf{D} (decode), and \textbf{R} (replica).
The routing parameter $x \in [0,1]$ determines D's Turn 2+ behavior: when $x{=}0$, D receives KV transfer from P every turn (traditional PD); when $x{=}1$, D processes new input tokens locally using cached prefix (Full AP-to-D), eliminating network transfer overhead.
\Cref{fig:configuration}a illustrates how these routing schemes operate: Replica (local execution), PD ($x{=}0$: P$\to$D with KV transfer every turn), and Full AP-to-D ($x{=}1$: P$\to$D with KV transfer only on Turn 1).

\paragraph{Configuration Space.}
Using these three machine types with 4 GPUs, we explore 17 configurations, including 7 hybrid configurations that combine R with P/D (see \Cref{tab:all-configs} in Appendix).
However, since hybrid configurations generally underperform pure PD alternatives and are orthogonal to AP routing optimization, we concentrate on 10 core configurations:
\textbf{Replica} (1 config: all local execution),
$\bm{x=0}$ configurations (3 configs: traditional PD with KV transfer every turn),
$\bm{x=1}$ configurations (3 configs: Full AP-to-D with local Turn 2+ processing),
and $\bm{0<x<1}$ (3 configs: partial AP routing to D, e.g., $x{=}\frac{1}{3},\frac{1}{2},\frac{2}{3}$).

\paragraph{Workloads.}
We design 18 synthetic workloads by combining 2 Turn 1 input/output length settings with 9 Turn 2 settings.
Each setting specifies a fixed $(n_{\text{in}}, n_{\text{out}})$ pair, the number of input and output tokens for that turn.
The 9 Turn 2 settings span three categories: \textit{decode-heavy} (4 settings with short input and long output, e.g., $n_{\text{in}}{=}32, n_{\text{out}}{=}512$), \textit{balanced} (2 settings with similar lengths), and \textit{prefill-heavy} (3 settings with long input and short output, e.g., $n_{\text{in}}{=}1024, n_{\text{out}}{=}32$), covering the full spectrum of multi-turn conversation patterns.

\paragraph{Experimental Setup.}
All experiments were run on an internal cluster node with 4 NVIDIA H100 80GB HBM3 GPUs connected via NVLink. We evaluate the impact of slower inter-node interconnects via bandwidth simulation in \Cref{subsec:scaling-sim}.
We use Llama-3.1-8B as the primary model, with validation experiments on Qwen2.5-14B and Qwen3-30B-A3B.
Each configuration is evaluated at 10 QPS (queries per second) levels (0.5, 1, 2, 4, 6, 8, 10, 12, 16, 20), yielding $17 \times 18 \times 10 = 3{,}060$ data points.
For each test point, we run a fixed 10-second duration with conversations arriving according to a Poisson process at the target QPS (e.g., QPS=4 generates $\sim$40 two-turn conversations).
We measure Turn 1 TTFT, Turn 2 TTFT, average TPOT, throughput (tokens/sec), and success rate.

\subsection{Full AP-to-D Advantage: Eliminating KV Transfer Overhead}
\label{subsec:ppd-advantage}

We now quantify the advantage of $x{=}1$ configurations (Full AP-to-D) over traditional PD disaggregation ($x{=}0$).
The core finding: \textbf{switching from $x{=}0$ to $x{=}1$ reduces Turn 2 TTFT by 48--73\%} by eliminating KV transfer for follow-up turns.
\Cref{tab:ppd-improvement} compares matched configurations: $x{=}1$ consistently outperforms $x{=}0$ on Turn 2 TTFT, with 1P\_3D achieving up to 73.3\% improvement and even 3P\_1D showing 24.9--44.3\% gains.

\begin{table}[htbp]
\caption{Turn 2 TTFT improvement when switching from $x{=}0$ to $x{=}1$. Negative values indicate improvement. The $x{=}1$ advantage increases with load for P-scarce configurations (1P, 2P) but diminishes when P is abundant (3P).}
\label{tab:ppd-improvement}
\centering
\small
\begin{tabular}{lccc}
\toprule
\textbf{Config} & \textbf{Low QPS} & \textbf{Med QPS} & \textbf{High QPS} \\
& (0.5--2) & (4--8) & (12--20) \\
\midrule
1P\_3D: $x{=}0 \to x{=}1$ & $-$57.8\% & $-$65.2\% & $\mathbf{-73.3\%}$ \\
2P\_2D: $x{=}0 \to x{=}1$ & $-$47.7\% & $-$51.6\% & $-$56.2\% \\
3P\_1D: $x{=}0 \to x{=}1$ & $-$44.3\% & $-$38.1\% & $-$24.9\% \\
\bottomrule
\end{tabular}
\end{table}

A striking pattern emerges: \textit{the $x{=}1$ advantage increases with load}, since as QPS rises, transfer queuing grows and D's local cache access becomes increasingly valuable.
The table also reveals that the $x{=}1$ advantage is maximized when P resources are scarce: 1P shows up to 73.3\% improvement while 3P shows diminishing returns.
In other words, \textbf{the fewer prefill nodes available, the greater the benefit of local processing on D}, because P becomes a bottleneck that $x{=}1$ bypasses entirely.

When P nodes are scarce, they become a bottleneck. In $x{=}0$ mode, Turn 2+ must traverse this bottleneck (P$\to$D), while $x{=}1$ bypasses P entirely.
This makes $x{=}1$ configurations more resilient: at high QPS, they exhibit lower failure rates (see \Cref{tab:failure-rates} in Appendix) since only Turn 1 contends for P capacity.
When P is abundant, prefill capacity is no longer the bottleneck, and local AP overhead becomes relatively more significant.

\paragraph{Validation: Turn and Model Scaling.}
The $x{=}1$ advantage generalizes beyond 2-turn conversations and the primary 8B model.
Scaling experiments (see \Cref{fig:validation} in Appendix) show stable $\sim$70\% Turn 2+ TTFT improvement across 2--16 turns and three model sizes (8B, 14B, 30B), confirming that the benefit stems from architectural properties rather than model-specific characteristics.

\subsection{No Universal Best: Objective-Dependent Optimization}
\label{subsec:no-universal-best}

The previous section established $x{=}1$'s clear advantage for Turn 2 TTFT.
However, production LLM serving must balance multiple objectives: latency (TTFT), generation smoothness (TPOT), and system efficiency (throughput).
When we expand our analysis to these dimensions, a more nuanced picture emerges.

\paragraph{Core Finding.}
\textit{92.2\% of workload-QPS combinations have different optimal configurations for Turn 2 TTFT versus Avg TPOT.}
This fundamental tension means no single configuration dominates across all objectives. The optimal choice depends on the metric being optimized.

\Cref{tab:winner-distribution} quantifies this trade-off by showing the winner percentage for each configuration category across three objectives.
Three patterns stand out:

\begin{table}[t]
\caption{Winner distribution across three optimization objectives: minimizing Turn 2 TTFT, minimizing TPOT, and maximizing throughput. Each cell shows the percentage of (workload, QPS) combinations where that configuration category achieves the best result. Columns do not sum to 100\% as hybrid configurations (which rarely win) are excluded. Full AP-to-D ($x{=}1$) configurations demonstrate the most balanced competitiveness, achieving the highest average win rate across all objectives.}
\label{tab:winner-distribution}
\centering
\small
\begin{tabular}{lcccc}
\toprule
\textbf{Mode} & \textbf{TTFT} & \textbf{TPOT} & \textbf{Thpt} & \textbf{Avg} \\
\midrule
Replica (4R) & \textbf{63.3\%} & 0.6\% & 0\% & 21.3\% \\
\midrule
$x{=}0$ (PD) & 0\% & \textbf{38.3\%} & 4.4\% & 14.2\% \\
$0{<}x{<}1$ & 3.3\% & 33.3\% & 27.8\% & 21.5\% \\
\rowcolor{gray!15} $x{=}1$ (Full AP-to-D) & 27.2\% & 15.6\% & \textbf{38.3\%} & \textbf{27.0\%} \\
\bottomrule
\end{tabular}
\end{table}

\textit{(1) Replica dominates Turn 2 TTFT}.
With zero network transfer and local prefix caching, Replica achieves the lowest Turn 2 latency.
However, Replica wins almost no TPOT or throughput scenarios: its lack of workload isolation lets prefill operations interfere with decode batches.

\textit{(2) $x{=}0$ and $0{<}x{<}1$ configurations dominate TPOT}.
Physical separation of prefill and decode workloads ensures stable token generation without interference.
The decode-only (D) machines maintain consistent batch sizes and dedicated memory bandwidth, yielding predictable per-token latency.

\textit{(3) Full AP-to-D ($x{=}1$) is the best disaggregated option for Turn 2 TTFT}.
Among disaggregated configurations, $x{=}1$ captures over a quarter of Turn 2 TTFT wins while $x{=}0$ captures none.
\textbf{$x{=}1$ also leads to throughput dominance} (38.3\%): its cache reuse and reduced network load translate to higher system efficiency.

\paragraph{Full AP-to-D's Balanced Profile.}
A key observation from \Cref{tab:winner-distribution}: $x{=}1$ (Full AP-to-D) is the \textit{only} disaggregated mode that is competitive across all three objectives.
While $x{=}0$ and $0{<}x{<}1$ configurations specialize in TPOT optimization, they rarely win on Turn 2 TTFT.
$x{=}1$ achieves the highest average win rate across objectives (see Avg column), making it a robust default choice when workload characteristics are uncertain. Production deployments prioritizing smooth token delivery may still prefer $x{=}0$ configurations, which PPD recovers as the extreme weight setting (\Cref{subsec:weight-tradeoff}).

\section{PPD: Dynamic AP Routing System}
\label{sec:ppd-design}

The trade-off analysis in \Cref{subsec:no-universal-best} reveals that no single static $x$ dominates all objectives. We therefore design PPD, a dynamic routing system that selects $x$ per request from workload characteristics and operator weights, following workload-aware scheduling principles~\cite{fu2024efficientllmschedulinglearning, jain2025intelligentrouterllmworkloads}.

\begin{algorithm}[t]
\caption{PPD Dynamic Routing}
\label{alg:routing}
\begin{algorithmic}[1]
    \State \textbf{Phase 1: Offline Table Construction}
    \For{each discretized workload $\hat{\psi}$ in benchmark grid}
        \State Measure Turn 2 TTFT and TPOT at $x{=}0$ and $x{=}1$
        \State Compute $S(\hat{\psi};\, \pi, \mathbf{w})$ via \Cref{eq:score}
        \State Store $x^*(\hat{\psi}) \gets \mathbb{1}[S > 0]$
    \EndFor
    \State
    \State \textbf{Phase 2: Online Per-Request Decision}
    \State \textbf{Input:} request workload $\psi$ at turn $t$
    \If{$t = 1$} \Return $x = 0$ \Comment{no cached KV yet}
    \EndIf
    \State Discretize $\psi$ to nearest grid entry $\hat{\psi}$
    \State \Return $x^*(\hat{\psi})$
\end{algorithmic}
\end{algorithm}

PPD operates in two phases (\Cref{alg:routing}). Offline, it builds a lookup table over a coarse workload grid by directly measuring TTFT and TPOT at the two extremes $x{=}0$ and $x{=}1$ for each cell, then storing the sign-of-score decision. Online, each Turn 2+ request is mapped to the nearest cell along three axes (accumulated context length, input/output ratio, system QPS) and the precomputed decision is returned in $<$1\,ms; Turn 1 always returns $x{=}0$ since no cached context exists. Discretization thresholds and the request feature schema appear in \Cref{app:discretization}.

\begin{figure*}[t]
    \centering
    \includegraphics[width=\textwidth]{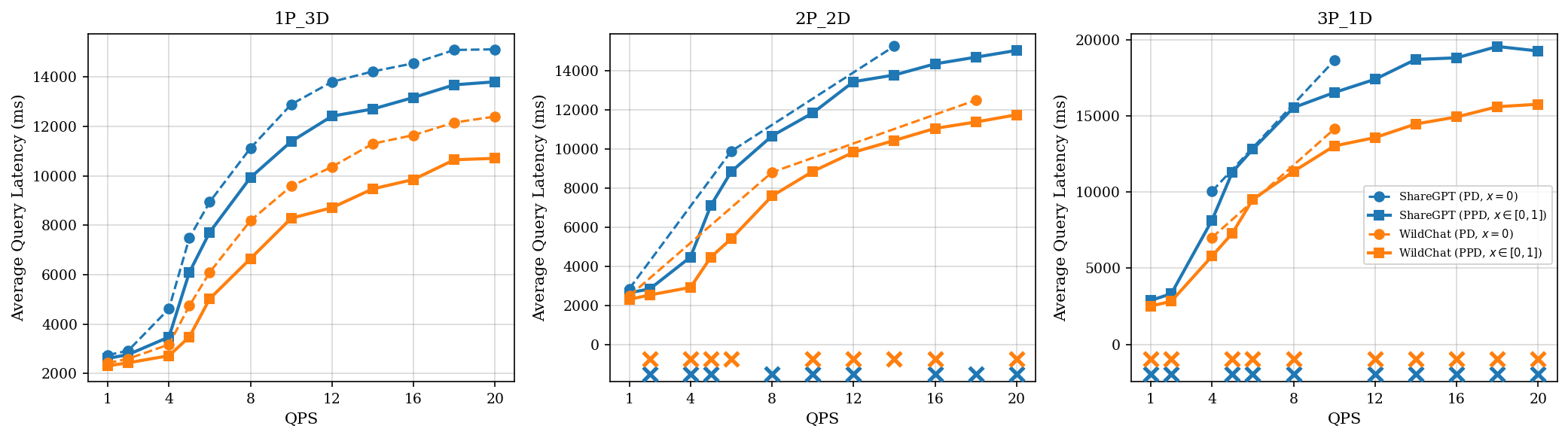}
    \caption{
        \textbf{PPD improves stability and reduces latency.}
        Average query latency vs.\ QPS for three configurations (1P\_3D, 2P\_2D, 3P\_1D) on ShareGPT (blue) and WildChat (orange) datasets.
        Dashed lines: PD ($x{=}0$); solid lines: PPD ($x \in [0,1]$).
        \textbf{$\times$ markers indicate service degradation} (success rate $<$95\% due to request timeout).
        PPD consistently achieves lower latency than PD while maintaining stability across the entire QPS range.
    }
    \label{fig:real-validation}
\end{figure*}

\paragraph{Decoupling Configuration Sizing from Turn 2+ SLO Tuning.}
PPD splits the multi-turn serving control problem into two independent knobs. In traditional PD, the P:D ratio simultaneously dictates Turn 1 prefill capacity and Turn 2+ latency trade-offs, forcing the operator to balance a coupled multi-objective parameter. PPD decouples these: the P:D ratio governs Turn 1 throughput, while the operator weights $\mathbf{w}$ select the Turn 2+ operating point on the Pareto frontier. Empirically, PPD remains stable across all three P:D ratios we evaluate (\Cref{fig:real-validation}), whereas the static $x{=}0$ baseline collapses when P is scarce.

Our prototype builds on vLLM's disaggregated serving infrastructure, reusing its KV transfer protocol and prefix cache. Implementation details, including session management and routing protocol, are provided in \Cref{app:implementation}.

\section{Real-World Validation}
\label{sec:validation}

Synthetic workloads provide well-controlled conditions for systematically exploring the configuration space and isolating individual factors.
To validate that these findings generalize to the heterogeneous distributions found in real traffic, we further evaluate PPD on real-world multi-turn conversation datasets.

\subsection{Datasets and Setup}
\label{subsec:datasets}

We use two publicly available multi-turn conversation datasets:
\textbf{ShareGPT}~\cite{vicuna2023} with user-shared ChatGPT conversations with diverse topics and interaction styles, and \textbf{WildChat}~\cite{zhao2024wildchat1mchatgptinteraction} with in-the-wild user conversations with varied prefill-to-decode ratios.

We filter multi-turn conversations from both datasets and evaluate three representative PD configurations (1P\_3D, 2P\_2D, 3P\_1D) with and without dynamic PPD enabled at QPS levels from 1 to 20.
The dynamic PPD mode uses balanced weights ($w_{\text{ttft}} = w_{\text{tpot}} = 1.0$).
We measure Turn 2+ TTFT, average query latency, throughput, and success rate (defined as $\geq$95\% of requests completing without timeout).

\subsection{PPD Improves Both Stability and Latency}
\label{subsec:kv-bottleneck}

Our real-world experiments (\Cref{fig:real-validation}) reveal two key findings: (1) PPD achieves consistently lower average query latency than baseline ($x{=}0$) across both datasets, and (2) PPD resolves the KV transfer bottleneck that causes $x{=}0$ mode degradation under multi-turn workloads.
A complete three-way comparison including the $x{=}1$ static baseline is provided in \Cref{app:3way}.

\paragraph{Finding 1: Lower Latency.}
As shown in \Cref{fig:real-validation}, PPD curves (solid lines) consistently lie below baseline curves (dashed lines) across all configurations and QPS levels where both modes succeed.
For the stable 1P\_3D configuration on ShareGPT, PPD reduces average query latency by 15--25\% across the QPS range.
This improvement stems from PPD's elimination of redundant KV transfers for Turn 2+ requests: instead of traversing the P$\to$D network path every turn, follow-up requests execute locally on decode nodes using cached prefixes.

\paragraph{Finding 2: Resolving Service Instability.}
Beyond latency improvement, PPD transforms previously unusable configurations into stable deployments.
The baseline ($x{=}0$) configurations exhibit service degradation (defined as success rate $<$95\%, primarily caused by request queuing and timeout under KV transfer saturation, not memory exhaustion or hardware failure):
2P\_2D and 3P\_1D's $x{=}0$ baselines degrade across the majority of QPS levels in both datasets, while only 1P\_3D remains stable where sufficient D capacity absorbs the KV load.

In contrast, \textit{PPD-enabled configurations achieve 100\% success rate across all QPS levels and both datasets}.
The $\times$ markers in \Cref{fig:real-validation} visually demonstrate this gap: baseline curves are fragmented with numerous failure points, while PPD curves remain complete and continuous.

\paragraph{Root Cause of Baseline Degradation.}
The root cause is \textit{KV transfer saturation}: in $x{=}0$ mode (traditional PD), decode nodes receive KV transfers for \textit{every} turn, while PPD dynamically routes Turn 2+ requests with higher $x$ values, cutting KV transfer load by ${\sim}75\%$ at the observed average of 3.1 turns per conversation.
At 3.1 average turns per conversation, this creates a $\sim$3$\times$ difference in network load, explaining why PD configurations degrade while PPD remains stable.

\subsection{Comparison with the Strongest Static Baseline (x=1)}
\label{subsec:x1-comparison}

We next isolate PPD against the strongest static disaggregated baseline, $x{=}1$ (Full AP-to-D), the simplest heuristic that keeps every Turn 2+ request local. On end-to-end average latency the two appear visually similar (\Cref{fig:3way-e2e} in Appendix), which is expected under balanced weights $w_{\text{ttft}}{=}w_{\text{tpot}}{=}1$: PPD trades some TTFT for TPOT improvement on requests it routes back to P, and these gains and losses cancel in the composite metric. A per-metric decomposition (\Cref{tab:metric-wins}) tells a different story.

\begin{table}[h]
\caption{Per-metric winner counts across 27 test points (3 configs $\times$ 9 QPS, WildChat). PPD achieves the most TPOT wins (surpassing $x{=}0$) and the most TTFT wins (surpassing $x{=}1$), while matching $x{=}1$'s 100\% success rate. Neither static baseline dominates all metrics; PPD is the only mode competitive across all three.}
\label{tab:metric-wins}
\centering
\small
\begin{tabular}{lccc}
\toprule
\textbf{Mode} & \textbf{TPOT Best} & \textbf{TTFT Best} & \textbf{SR = 100\%} \\
\midrule
$x{=}0$ & 10/27 & 0/27 & 4/27 \\
$x{=}1$ & 5/27 & 13/27 & 27/27 \\
\rowcolor{gray!15} \textbf{PPD} & \textbf{12/27} & \textbf{14/27} & \textbf{27/27} \\
\bottomrule
\end{tabular}
\end{table}

PPD is the only mode that is competitive across all three axes: it beats $x{=}0$ on TPOT (12 vs.\ 10 wins) and beats $x{=}1$ on TTFT (14 vs.\ 13 wins), while matching $x{=}1$'s 100\% success rate. Since $x{=}1$ is already the simplest static rule that exploits local KV at all, any weaker heuristic (e.g., random routing) is strictly dominated; the meaningful comparison is therefore whether PPD improves over $x{=}1$, which \Cref{tab:metric-wins} confirms it does on each metric individually. The weights $(w_{\text{ttft}}, w_{\text{tpot}})$ are precisely the control knob: PPD recovers $x{=}0$ or $x{=}1$ as special cases at extreme settings, and \Cref{subsec:weight-tradeoff} characterizes the smooth interpolation.

\subsection{Robustness Across Network Speeds}
\label{subsec:scaling-sim}

Our experiments use intra-node NVLink, but production multi-node deployments rely on RDMA over InfiniBand or Converged Ethernet (RoCE), typically 2--8$\times$ slower~\cite{patel2024splitwiseefficientgenerativellm, qin2025mooncakekvcachecentricdisaggregatedarchitecture}. Since PPD bypasses the P$\to$D channel for Turn 2+ requests, its advantage should grow as the network slows. To verify this without multi-node hardware, we follow the bandwidth-emulation methodology of TetriInfer~\cite{hu2024inferenceinterferencedisaggregatellm} and inject a calibrated extra delay on each PD-routed request, leaving PPD's local path untouched. The exact delay model and the per-token KV footprint are in \Cref{app:bandwidth-sim}.

\begin{figure*}[t]
    \centering
    \includegraphics[width=\textwidth]{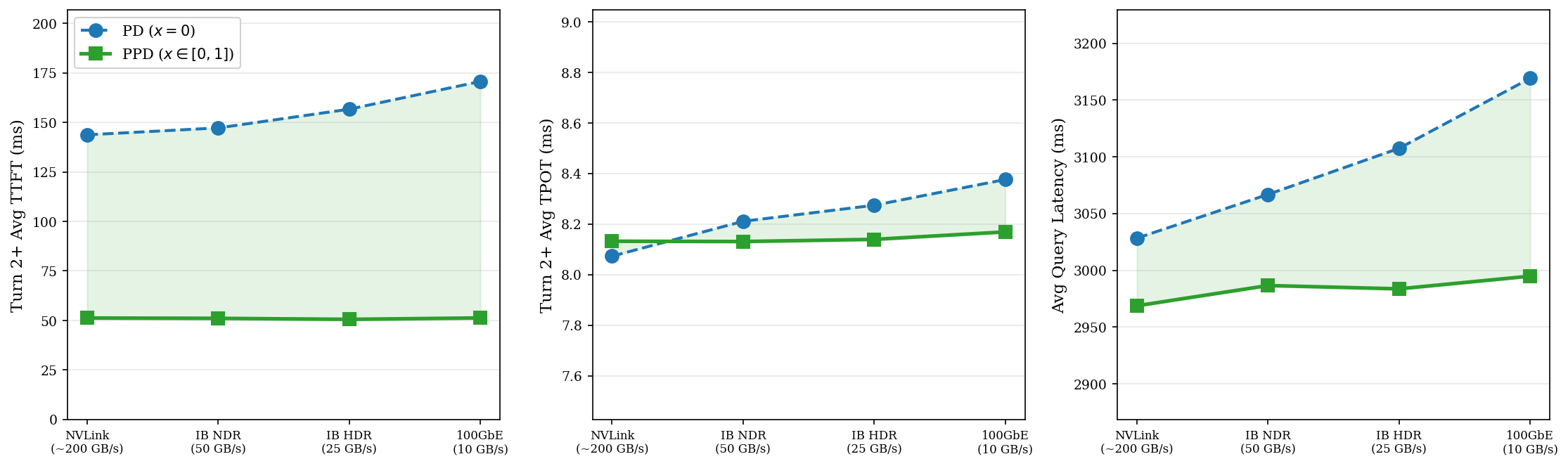}
    \caption{
        \textbf{PPD's advantage grows monotonically as the simulated network slows.}
        Results on 1P\_3D, QPS=1, WildChat (500 conversations) across four interconnects: NVLink ($\sim$150~GB/s effective, intra-node), InfiniBand NDR (50~GB/s), InfiniBand HDR (25~GB/s), and 100GbE (10~GB/s).
        \textbf{Left:} Turn 2+ TTFT. PD ($x{=}0$) rises from 143.7\,ms to 170.6\,ms ($+$18.7\%) as bandwidth drops; PPD remains flat at $\sim$51\,ms.
        \textbf{Center:} Turn 2+ TPOT. PD drifts from 8.07\,ms to 8.38\,ms ($+$3.8\%) under network contention; PPD stays near 8.1\,ms.
        \textbf{Right:} End-to-end latency. PD increases by 4.7\% (3{,}028\,ms $\to$ 3{,}169\,ms); PPD by 0.9\%, attributable solely to Turn 1.
    }
    \label{fig:scaling-sim}
\end{figure*}

\Cref{fig:scaling-sim} sweeps four interconnects from NVLink down to 100GbE. As effective bandwidth drops by an order of magnitude, the PD baseline's Turn 2+ TTFT rises by ${\sim}19\%$ while PPD stays at ${\sim}51$\,ms; the relative TTFT reduction widens from 64\% to 70\%, with TPOT and end-to-end latency tracking the same trend. NVLink-based evaluation is therefore a conservative lower bound for PPD's advantage in real multi-node deployments.

\subsection{Case Study: Weight-Based TTFT-TPOT Trade-Off}
\label{subsec:weight-tradeoff}

So far we ran PPD with balanced weights ($w_{\text{ttft}}{=}w_{\text{tpot}}{=}1$). To illustrate the operator-facing trade-off, we sweep $w_{\text{tpot}}$ on a slice of the workload where PPD is most challenged: 500 prefill-heavy multi-turn conversations sampled from ShareGPT and WildChat, run on 1P\_3D at QPS 8 and 16. As $w_{\text{tpot}}$ rises, the decision engine increasingly prefers the PD path over local processing.

\begin{figure}[t]
    \centering
    \includegraphics[width=\columnwidth]{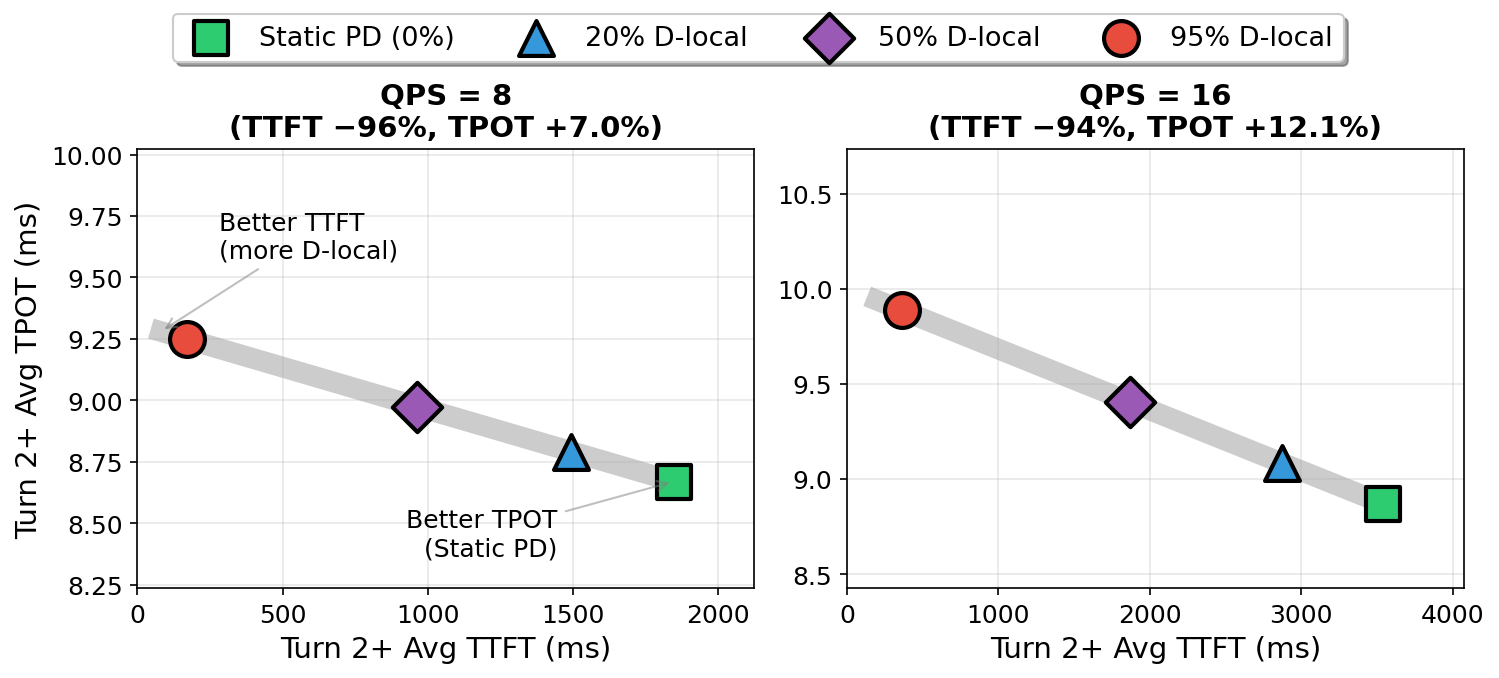}
    \caption{
        \textbf{$w_{\text{tpot}}$ traces a monotonic frontier between TTFT and TPOT.}
        Turn 2+ latency on 1P\_3D under prefill-heavy workloads. Operating points (left to right): static PD ($0\%$ D-local); $w_{\text{tpot}}{=}6$ ($20\%$); $w_{\text{tpot}}{=}3$ ($50\%$); balanced $w_{\text{tpot}}{=}1$ ($95\%$).
    }
    \label{fig:weight-tradeoff}
\end{figure}

\Cref{fig:weight-tradeoff} reveals that the weight ratio $w_{\text{tpot}}/w_{\text{ttft}}$ is a \emph{monotonic control surface}: it smoothly interpolates between the $x{=}1$ regime (94--96\% TTFT reduction, 7--12\% TPOT degradation) and the $x{=}0$ baseline. Operators thus pick a single point on this surface offline, rather than tuning multiple coupled parameters online. PPD is a \emph{routing actuator}: it does not enforce end-to-end SLO bounds (which require closed-loop admission control and batch scheduling), but exposes a predictable knob that higher-level SLO controllers can drive from observed P99 metrics.

\section{Discussion}
\label{sec:discuss}

\paragraph{Relationship with Concurrent Work and Limitations.}
Concurrently, AMPD~\cite{he2026efficientmultiroundllminference} identifies the same multi-turn inefficiency in PD disaggregation and proposes coordinating prefill workloads via real-time queue-state estimation paired with offline hardware planning. Our work is complementary in scope: PPD contributes (i) a micro-architectural account of \emph{why} local append-prefill is viable, namely the order-of-magnitude interference gap (\Cref{sec:interference}); (ii) a clean optimization formulation in which traditional PD is the special case $x \equiv 0$; and (iii) a single-knob TTFT--TPOT control surface (\Cref{eq:score}) that operators can tune offline. The principal limitation of PPD is that its lookup table is built offline and degrades gracefully but suboptimally when hardware or workload distributions drift far from the calibration set; AMPD's online queue-state estimation is a natural mechanism for closing this gap, and we view PPD's principled foundation and AMPD's adaptive loop as composable rather than competing. Other natural extensions include integration with higher-level SLO controllers that drive $\mathbf{w}$ from observed P99 metrics, deployment on heterogeneous GPU clusters~\cite{tong2025parallaxefficientllminference} or long-context elastic parallelism~\cite{wu2024loongserveefficientlyservinglongcontext}, and joint use with distributed KV cache layers (below).

\paragraph{Relationship with Distributed KV Cache.}
Distributed KV cache layers (e.g., Mooncake~\cite{qin2025mooncakekvcachecentricdisaggregatedarchitecture}, MemServe~\cite{hu2024memservecontextcachingdisaggregated}) operate at a different layer than PPD: they decide \emph{where} prefix states are stored across the cluster, while PPD decides \emph{how} cache-hitting and cache-missing requests are scheduled to avoid mutual interference. The two are fully compatible. Distributed storage maximizes prefix reuse, while PPD ensures that warm requests are not delayed by cold ones; we expect joint deployment to amplify both effects in large-scale serving.

\section{Conclusion}
\label{sec:conclusion}
Multi-turn conversations expose fundamental limitations in PD disaggregation that single-turn workloads do not reveal.
The order-of-magnitude interference gap between full prefill and append-prefill opens a new design dimension: decode nodes can safely handle Turn 2+ requests locally.
Yet our systematic exploration of 3,060 configurations confirms that static routing cannot universally optimize TTFT, TPOT, and throughput.
PPD realizes a workload-aware alternative through an offline-tuned, single-knob control surface (\Cref{eq:score}), cutting Turn 2+ TTFT by an average of $\sim$68\% on real-world workloads while keeping TPOT competitive and adding $<$1\,ms of per-request overhead.

\section*{Impact Statement}
This work improves the efficiency of multi-turn LLM serving, potentially reducing computational costs and energy consumption in large-scale deployments.
We do not foresee direct negative societal impacts beyond those inherent to LLM technology broadly.

\bibliography{main}

@misc{zhong2024distservedisaggregatingprefilldecoding,
      title={DistServe: Disaggregating Prefill and Decoding for Goodput-optimized Large Language Model Serving}, 
      author={Yinmin Zhong and Shengyu Liu and Junda Chen and Jianbo Hu and Yibo Zhu and Xuanzhe Liu and Xin Jin and Hao Zhang},
      year={2024},
      eprint={2401.09670},
      archivePrefix={arXiv},
      primaryClass={cs.DC},
      url={https://arxiv.org/abs/2401.09670}, 
}

@misc{patel2024splitwiseefficientgenerativellm,
      title={Splitwise: Efficient generative LLM inference using phase splitting}, 
      author={Pratyush Patel and Esha Choukse and Chaojie Zhang and Aashaka Shah and Íñigo Goiri and Saeed Maleki and Ricardo Bianchini},
      year={2024},
      eprint={2311.18677},
      archivePrefix={arXiv},
      primaryClass={cs.AR},
      url={https://arxiv.org/abs/2311.18677}, 
}

@misc{qin2025mooncakekvcachecentricdisaggregatedarchitecture,
      title={Mooncake: A KVCache-centric Disaggregated Architecture for LLM Serving}, 
      author={Ruoyu Qin and Zheming Li and Weiran He and Mingxing Zhang and Yongwei Wu and Weimin Zheng and Xinran Xu},
      year={2025},
      eprint={2407.00079},
      archivePrefix={arXiv},
      primaryClass={cs.DC},
      url={https://arxiv.org/abs/2407.00079}, 
}

@inproceedings {280922,
author = {Gyeong-In Yu and Joo Seong Jeong and Geon-Woo Kim and Soojeong Kim and Byung-Gon Chun},
title = {Orca: A Distributed Serving System for {Transformer-Based} Generative Models},
booktitle = {16th USENIX Symposium on Operating Systems Design and Implementation (OSDI 22)},
year = {2022},
isbn = {978-1-939133-28-1},
address = {Carlsbad, CA},
pages = {521--538},
url = {https://www.usenix.org/conference/osdi22/presentation/yu},
publisher = {USENIX Association},
month = jul
}

@misc{kwon2023efficientmemorymanagementlarge,
      title={Efficient Memory Management for Large Language Model Serving with PagedAttention}, 
      author={Woosuk Kwon and Zhuohan Li and Siyuan Zhuang and Ying Sheng and Lianmin Zheng and Cody Hao Yu and Joseph E. Gonzalez and Hao Zhang and Ion Stoica},
      year={2023},
      eprint={2309.06180},
      archivePrefix={arXiv},
      primaryClass={cs.LG},
      url={https://arxiv.org/abs/2309.06180}, 
}

@misc{zheng2024sglangefficientexecutionstructured,
      title={SGLang: Efficient Execution of Structured Language Model Programs}, 
      author={Lianmin Zheng and Liangsheng Yin and Zhiqiang Xie and Chuyue Sun and Jeff Huang and Cody Hao Yu and Shiyi Cao and Christos Kozyrakis and Ion Stoica and Joseph E. Gonzalez and Clark Barrett and Ying Sheng},
      year={2024},
      eprint={2312.07104},
      archivePrefix={arXiv},
      primaryClass={cs.AI},
      url={https://arxiv.org/abs/2312.07104}, 
}

@inproceedings {298679,
author = {Amey Agrawal and Nitin Kedia and Ashish Panwar and Jayashree Mohan and Nipun Kwatra and Bhargav Gulavani and Alexey Tumanov and Ramachandran Ramjee},
title = {Taming {Throughput-Latency} Tradeoff in {LLM} Inference with {Sarathi-Serve}},
booktitle = {18th USENIX Symposium on Operating Systems Design and Implementation (OSDI 24)},
year = {2024},
isbn = {978-1-939133-40-3},
address = {Santa Clara, CA},
pages = {117--134},
url = {https://www.usenix.org/conference/osdi24/presentation/agrawal},
publisher = {USENIX Association},
month = jul
}

@misc{deepseekai2025deepseekv3technicalreport,
      title={DeepSeek-V3 Technical Report}, 
      author={DeepSeek-AI and Aixin Liu and Bei Feng and Bing Xue and Bingxuan Wang and Bochao Wu and Chengda Lu and Chenggang Zhao and Chengqi Deng and Chenyu Zhang and Chong Ruan and Damai Dai and Daya Guo and Dejian Yang and Deli Chen and Dongjie Ji and Erhang Li and Fangyun Lin and Fucong Dai and Fuli Luo and Guangbo Hao and Guanting Chen and Guowei Li and H. Zhang and Han Bao and Hanwei Xu and Haocheng Wang and Haowei Zhang and Honghui Ding and Huajian Xin and Huazuo Gao and Hui Li and Hui Qu and J. L. Cai and Jian Liang and Jianzhong Guo and Jiaqi Ni and Jiashi Li and Jiawei Wang and Jin Chen and Jingchang Chen and Jingyang Yuan and Junjie Qiu and Junlong Li and Junxiao Song and Kai Dong and Kai Hu and Kaige Gao and Kang Guan and Kexin Huang and Kuai Yu and Lean Wang and Lecong Zhang and Lei Xu and Leyi Xia and Liang Zhao and Litong Wang and Liyue Zhang and Meng Li and Miaojun Wang and Mingchuan Zhang and Minghua Zhang and Minghui Tang and Mingming Li and Ning Tian and Panpan Huang and Peiyi Wang and Peng Zhang and Qiancheng Wang and Qihao Zhu and Qinyu Chen and Qiushi Du and R. J. Chen and R. L. Jin and Ruiqi Ge and Ruisong Zhang and Ruizhe Pan and Runji Wang and Runxin Xu and Ruoyu Zhang and Ruyi Chen and S. S. Li and Shanghao Lu and Shangyan Zhou and Shanhuang Chen and Shaoqing Wu and Shengfeng Ye and Shengfeng Ye and Shirong Ma and Shiyu Wang and Shuang Zhou and Shuiping Yu and Shunfeng Zhou and Shuting Pan and T. Wang and Tao Yun and Tian Pei and Tianyu Sun and W. L. Xiao and Wangding Zeng and Wanjia Zhao and Wei An and Wen Liu and Wenfeng Liang and Wenjun Gao and Wenqin Yu and Wentao Zhang and X. Q. Li and Xiangyue Jin and Xianzu Wang and Xiao Bi and Xiaodong Liu and Xiaohan Wang and Xiaojin Shen and Xiaokang Chen and Xiaokang Zhang and Xiaosha Chen and Xiaotao Nie and Xiaowen Sun and Xiaoxiang Wang and Xin Cheng and Xin Liu and Xin Xie and Xingchao Liu and Xingkai Yu and Xinnan Song and Xinxia Shan and Xinyi Zhou and Xinyu Yang and Xinyuan Li and Xuecheng Su and Xuheng Lin and Y. K. Li and Y. Q. Wang and Y. X. Wei and Y. X. Zhu and Yang Zhang and Yanhong Xu and Yanhong Xu and Yanping Huang and Yao Li and Yao Zhao and Yaofeng Sun and Yaohui Li and Yaohui Wang and Yi Yu and Yi Zheng and Yichao Zhang and Yifan Shi and Yiliang Xiong and Ying He and Ying Tang and Yishi Piao and Yisong Wang and Yixuan Tan and Yiyang Ma and Yiyuan Liu and Yongqiang Guo and Yu Wu and Yuan Ou and Yuchen Zhu and Yuduan Wang and Yue Gong and Yuheng Zou and Yujia He and Yukun Zha and Yunfan Xiong and Yunxian Ma and Yuting Yan and Yuxiang Luo and Yuxiang You and Yuxuan Liu and Yuyang Zhou and Z. F. Wu and Z. Z. Ren and Zehui Ren and Zhangli Sha and Zhe Fu and Zhean Xu and Zhen Huang and Zhen Zhang and Zhenda Xie and Zhengyan Zhang and Zhewen Hao and Zhibin Gou and Zhicheng Ma and Zhigang Yan and Zhihong Shao and Zhipeng Xu and Zhiyu Wu and Zhongyu Zhang and Zhuoshu Li and Zihui Gu and Zijia Zhu and Zijun Liu and Zilin Li and Ziwei Xie and Ziyang Song and Ziyi Gao and Zizheng Pan},
      year={2025},
      eprint={2412.19437},
      archivePrefix={arXiv},
      primaryClass={cs.CL},
      url={https://arxiv.org/abs/2412.19437}, 
}

@misc{gao2024costefficientlargelanguagemodel,
      title={Cost-Efficient Large Language Model Serving for Multi-turn Conversations with CachedAttention}, 
      author={Bin Gao and Zhuomin He and Puru Sharma and Qingxuan Kang and Djordje Jevdjic and Junbo Deng and Xingkun Yang and Zhou Yu and Pengfei Zuo},
      year={2024},
      eprint={2403.19708},
      archivePrefix={arXiv},
      primaryClass={cs.CL},
      url={https://arxiv.org/abs/2403.19708}, 
}

@misc{liu2025lmcacheefficientkvcache,
      title={LMCache: An Efficient KV Cache Layer for Enterprise-Scale LLM Inference}, 
      author={Yuhan Liu and Yihua Cheng and Jiayi Yao and Yuwei An and Xiaokun Chen and Shaoting Feng and Yuyang Huang and Samuel Shen and Rui Zhang and Kuntai Du and Junchen Jiang},
      year={2025},
      eprint={2510.09665},
      archivePrefix={arXiv},
      primaryClass={cs.LG},
      url={https://arxiv.org/abs/2510.09665}, 
}

@misc{hu2024memservecontextcachingdisaggregated,
      title={MemServe: Context Caching for Disaggregated LLM Serving with Elastic Memory Pool}, 
      author={Cunchen Hu and Heyang Huang and Junhao Hu and Jiang Xu and Xusheng Chen and Tao Xie and Chenxi Wang and Sa Wang and Yungang Bao and Ninghui Sun and Yizhou Shan},
      year={2024},
      eprint={2406.17565},
      archivePrefix={arXiv},
      primaryClass={cs.DC},
      url={https://arxiv.org/abs/2406.17565}, 
}

@online{nvidia2025dynamo,
  title={{NVIDIA Dynamo}: A Low-Latency Distributed Inference Framework
         for Scaling Reasoning {AI} Models},
  author={{NVIDIA}},
  year={2025},
  url={https://developer.nvidia.com/dynamo},
  urldate={2025-01-29}
}

@misc{vicuna2023,
    title = {Vicuna: An Open-Source Chatbot Impressing GPT-4 with 90\%* ChatGPT Quality},
    url = {https://lmsys.org/blog/2023-03-30-vicuna/},
    author = {Chiang, Wei-Lin and Li, Zhuohan and Lin, Zi and Sheng, Ying and Wu, Zhanghao and Zhang, Hao and Zheng, Lianmin and Zhuang, Siyuan and Zhuang, Yonghao and Gonzalez, Joseph E. and Stoica, Ion and Xing, Eric P.},
    month = {March},
    year = {2023}
}

@misc{zhao2024wildchat1mchatgptinteraction,
      title={WildChat: 1M ChatGPT Interaction Logs in the Wild}, 
      author={Wenting Zhao and Xiang Ren and Jack Hessel and Claire Cardie and Yejin Choi and Yuntian Deng},
      year={2024},
      eprint={2405.01470},
      archivePrefix={arXiv},
      primaryClass={cs.CL},
      url={https://arxiv.org/abs/2405.01470}, 
}

@misc{yao2025cacheblendfastlargelanguage,
      title={CacheBlend: Fast Large Language Model Serving for RAG with Cached Knowledge Fusion}, 
      author={Jiayi Yao and Hanchen Li and Yuhan Liu and Siddhant Ray and Yihua Cheng and Qizheng Zhang and Kuntai Du and Shan Lu and Junchen Jiang},
      year={2025},
      eprint={2405.16444},
      archivePrefix={arXiv},
      primaryClass={cs.LG},
      url={https://arxiv.org/abs/2405.16444}, 
}

@misc{chen2024magicpiglshsamplingefficient,
      title={MagicPIG: LSH Sampling for Efficient LLM Generation}, 
      author={Zhuoming Chen and Ranajoy Sadhukhan and Zihao Ye and Yang Zhou and Jianyu Zhang and Niklas Nolte and Yuandong Tian and Matthijs Douze and Leon Bottou and Zhihao Jia and Beidi Chen},
      year={2024},
      eprint={2410.16179},
      archivePrefix={arXiv},
      primaryClass={cs.CL},
      url={https://arxiv.org/abs/2410.16179}, 
}

@misc{sun2025shadowkvkvcacheshadows,
      title={ShadowKV: KV Cache in Shadows for High-Throughput Long-Context LLM Inference}, 
      author={Hanshi Sun and Li-Wen Chang and Wenlei Bao and Size Zheng and Ningxin Zheng and Xin Liu and Harry Dong and Yuejie Chi and Beidi Chen},
      year={2025},
      eprint={2410.21465},
      archivePrefix={arXiv},
      primaryClass={cs.LG},
      url={https://arxiv.org/abs/2410.21465}, 
}

@misc{liu2025speculativeprefillturbochargingttft,
      title={Speculative Prefill: Turbocharging TTFT with Lightweight and Training-Free Token Importance Estimation}, 
      author={Jingyu Liu and Beidi Chen and Ce Zhang},
      year={2025},
      eprint={2502.02789},
      archivePrefix={arXiv},
      primaryClass={cs.CL},
      url={https://arxiv.org/abs/2502.02789}, 
}

@misc{shi2024discoveringgemsearlylayers,
      title={Discovering the Gems in Early Layers: Accelerating Long-Context LLMs with 1000x Input Token Reduction}, 
      author={Zhenmei Shi and Yifei Ming and Xuan-Phi Nguyen and Yingyu Liang and Shafiq Joty},
      year={2024},
      eprint={2409.17422},
      archivePrefix={arXiv},
      primaryClass={cs.CL},
      url={https://arxiv.org/abs/2409.17422}, 
}

@misc{tong2025parallaxefficientllminference,
      title={Parallax: Efficient LLM Inference Service over Decentralized Environment}, 
      author={Chris Tong and Youhe Jiang and Gufeng Chen and Tianyi Zhao and Sibian Lu and Wenjie Qu and Eric Yang and Lynn Ai and Binhang Yuan},
      year={2025},
      eprint={2509.26182},
      archivePrefix={arXiv},
      primaryClass={cs.DC},
      url={https://arxiv.org/abs/2509.26182}, 
}

@misc{gao2025duetserveharmonizingprefilldecode,
      title={DuetServe: Harmonizing Prefill and Decode for LLM Serving via Adaptive GPU Multiplexing}, 
      author={Lei Gao and Chaoyi Jiang and Hossein Entezari Zarch and Daniel Wong and Murali Annavaram},
      year={2025},
      eprint={2511.04791},
      archivePrefix={arXiv},
      primaryClass={cs.LG},
      url={https://arxiv.org/abs/2511.04791}, 
}

@misc{wang2025prefilldecodeaggregationdisaggregationunifying,
      title={Prefill-Decode Aggregation or Disaggregation? Unifying Both for Goodput-Optimized LLM Serving}, 
      author={Chao Wang and Pengfei Zuo and Zhangyu Chen and Yunkai Liang and Zhou Yu and Ming-Chang Yang},
      year={2025},
      eprint={2508.01989},
      archivePrefix={arXiv},
      primaryClass={cs.DC},
      url={https://arxiv.org/abs/2508.01989}, 
}

@misc{shi2025nexusproactiveintragpudisaggregationprefill,
      title={Nexus:Proactive Intra-GPU Disaggregation of Prefill and Decode in LLM Serving}, 
      author={Xiaoxiang Shi and Colin Cai and Junjia Du and Zhihao Jia},
      year={2025},
      eprint={2507.06608},
      archivePrefix={arXiv},
      primaryClass={cs.DC},
      url={https://arxiv.org/abs/2507.06608}, 
}

@misc{sun2024llumnixdynamicschedulinglarge,
      title={Llumnix: Dynamic Scheduling for Large Language Model Serving}, 
      author={Biao Sun and Ziming Huang and Hanyu Zhao and Wencong Xiao and Xinyi Zhang and Yong Li and Wei Lin},
      year={2024},
      eprint={2406.03243},
      archivePrefix={arXiv},
      primaryClass={cs.AR},
      url={https://arxiv.org/abs/2406.03243}, 
}

@misc{dao2022flashattentionfastmemoryefficientexact,
      title={FlashAttention: Fast and Memory-Efficient Exact Attention with IO-Awareness}, 
      author={Tri Dao and Daniel Y. Fu and Stefano Ermon and Atri Rudra and Christopher Ré},
      year={2022},
      eprint={2205.14135},
      archivePrefix={arXiv},
      primaryClass={cs.LG},
      url={https://arxiv.org/abs/2205.14135}, 
}

@misc{shah2024flashattention3fastaccurateattention,
      title={FlashAttention-3: Fast and Accurate Attention with Asynchrony and Low-precision}, 
      author={Jay Shah and Ganesh Bikshandi and Ying Zhang and Vijay Thakkar and Pradeep Ramani and Tri Dao},
      year={2024},
      eprint={2407.08608},
      archivePrefix={arXiv},
      primaryClass={cs.LG},
      url={https://arxiv.org/abs/2407.08608}, 
}

@misc{duan2023botchatevaluatingllmscapabilities,
      title={BotChat: Evaluating LLMs' Capabilities of Having Multi-Turn Dialogues}, 
      author={Haodong Duan and Jueqi Wei and Chonghua Wang and Hongwei Liu and Yixiao Fang and Songyang Zhang and Dahua Lin and Kai Chen},
      year={2023},
      eprint={2310.13650},
      archivePrefix={arXiv},
      primaryClass={cs.CL},
      url={https://arxiv.org/abs/2310.13650}, 
}

@misc{li2024snapkvllmknowslooking,
      title={SnapKV: LLM Knows What You are Looking for Before Generation}, 
      author={Yuhong Li and Yingbing Huang and Bowen Yang and Bharat Venkitesh and Acyr Locatelli and Hanchen Ye and Tianle Cai and Patrick Lewis and Deming Chen},
      year={2024},
      eprint={2404.14469},
      archivePrefix={arXiv},
      primaryClass={cs.CL},
      url={https://arxiv.org/abs/2404.14469}, 
}

@misc{liu2024cachegenkvcachecompression,
      title={CacheGen: KV Cache Compression and Streaming for Fast Large Language Model Serving}, 
      author={Yuhan Liu and Hanchen Li and Yihua Cheng and Siddhant Ray and Yuyang Huang and Qizheng Zhang and Kuntai Du and Jiayi Yao and Shan Lu and Ganesh Ananthanarayanan and Michael Maire and Henry Hoffmann and Ari Holtzman and Junchen Jiang},
      year={2024},
      eprint={2310.07240},
      archivePrefix={arXiv},
      primaryClass={cs.NI},
      url={https://arxiv.org/abs/2310.07240}, 
}

@misc{liu2025hamburgeracceleratingllminference,
      title={HAMburger: Accelerating LLM Inference via Token Smashing}, 
      author={Jingyu Liu and Ce Zhang},
      year={2025},
      eprint={2505.20438},
      archivePrefix={arXiv},
      primaryClass={cs.CL},
      url={https://arxiv.org/abs/2505.20438}, 
}

@misc{ye2024chunkattentionefficientselfattentionprefixaware,
      title={ChunkAttention: Efficient Self-Attention with Prefix-Aware KV Cache and Two-Phase Partition}, 
      author={Lu Ye and Ze Tao and Yong Huang and Yang Li},
      year={2024},
      eprint={2402.15220},
      archivePrefix={arXiv},
      primaryClass={cs.LG},
      url={https://arxiv.org/abs/2402.15220}, 
}

@misc{fu2024efficientllmschedulinglearning,
      title={Efficient LLM Scheduling by Learning to Rank}, 
      author={Yichao Fu and Siqi Zhu and Runlong Su and Aurick Qiao and Ion Stoica and Hao Zhang},
      year={2024},
      eprint={2408.15792},
      archivePrefix={arXiv},
      primaryClass={cs.LG},
      url={https://arxiv.org/abs/2408.15792}, 
}

@misc{2023lmdeploy,
    title={LMDeploy: A Toolkit for Compressing, Deploying, and Serving LLM},
    author={LMDeploy Contributors},
    howpublished = {\url{https://github.com/InternLM/lmdeploy}},
    year={2023}
}

@misc{ainslie2023gqatraininggeneralizedmultiquery,
      title={GQA: Training Generalized Multi-Query Transformer Models from Multi-Head Checkpoints}, 
      author={Joshua Ainslie and James Lee-Thorp and Michiel de Jong and Yury Zemlyanskiy and Federico Lebrón and Sumit Sanghai},
      year={2023},
      eprint={2305.13245},
      archivePrefix={arXiv},
      primaryClass={cs.CL},
      url={https://arxiv.org/abs/2305.13245}, 
}

@misc{munkhdalai2024leavecontextbehindefficient,
      title={Leave No Context Behind: Efficient Infinite Context Transformers with Infini-attention}, 
      author={Tsendsuren Munkhdalai and Manaal Faruqui and Siddharth Gopal},
      year={2024},
      eprint={2404.07143},
      archivePrefix={arXiv},
      primaryClass={cs.CL},
      url={https://arxiv.org/abs/2404.07143}, 
}

@misc{holmes2024deepspeedfastgenhighthroughputtextgeneration,
      title={DeepSpeed-FastGen: High-throughput Text Generation for LLMs via MII and DeepSpeed-Inference}, 
      author={Connor Holmes and Masahiro Tanaka and Michael Wyatt and Ammar Ahmad Awan and Jeff Rasley and Samyam Rajbhandari and Reza Yazdani Aminabadi and Heyang Qin and Arash Bakhtiari and Lev Kurilenko and Yuxiong He},
      year={2024},
      eprint={2401.08671},
      archivePrefix={arXiv},
      primaryClass={cs.PF},
      url={https://arxiv.org/abs/2401.08671}, 
}

@misc{wu2024loongserveefficientlyservinglongcontext,
      title={LoongServe: Efficiently Serving Long-Context Large Language Models with Elastic Sequence Parallelism}, 
      author={Bingyang Wu and Shengyu Liu and Yinmin Zhong and Peng Sun and Xuanzhe Liu and Xin Jin},
      year={2024},
      eprint={2404.09526},
      archivePrefix={arXiv},
      primaryClass={cs.DC},
      url={https://arxiv.org/abs/2404.09526}, 
}

@article{10.1145/3774909,
author = {Jeong, Jinwoo and Ahn, Jeongseob},
title = {An Efficient DNN Model Serving System using Layer-wise Caching and Direct-Host-Access},
year = {2026},
issue_date = {February 2026},
publisher = {Association for Computing Machinery},
address = {New York, NY, USA},
volume = {44},
number = {1},
issn = {0734-2071},
url = {https://doi.org/10.1145/3774909},
doi = {10.1145/3774909},
journal = {ACM Trans. Comput. Syst.},
month = jan,
articleno = {5},
numpages = {21}
}

@misc{yuan2024llminferenceunveiledsurvey,
      title={LLM Inference Unveiled: Survey and Roofline Model Insights}, 
      author={Zhihang Yuan and Yuzhang Shang and Yang Zhou and Zhen Dong and Zhe Zhou and Chenhao Xue and Bingzhe Wu and Zhikai Li and Qingyi Gu and Yong Jae Lee and Yan Yan and Beidi Chen and Guangyu Sun and Kurt Keutzer},
      year={2024},
      eprint={2402.16363},
      archivePrefix={arXiv},
      primaryClass={cs.CL},
      url={https://arxiv.org/abs/2402.16363}, 
}

@misc{he2026efficientmultiroundllminference,
      title={Efficient Multi-round LLM Inference over Disaggregated Serving}, 
      author={Wenhao He and Youhe Jiang and Penghao Zhao and Quanqing Xu and Eiko Yoneki and Bin Cui and Fangcheng Fu},
      year={2026},
      eprint={2602.14516},
      archivePrefix={arXiv},
      primaryClass={cs.DC},
      url={https://arxiv.org/abs/2602.14516}, 
}

@misc{hu2024inferenceinterferencedisaggregatellm,
      title={Inference without Interference: Disaggregate LLM Inference for Mixed Downstream Workloads}, 
      author={Cunchen Hu and Heyang Huang and Liangliang Xu and Xusheng Chen and Jiang Xu and Shuang Chen and Hao Feng and Chenxi Wang and Sa Wang and Yungang Bao and Ninghui Sun and Yizhou Shan},
      year={2024},
      eprint={2401.11181},
      archivePrefix={arXiv},
      primaryClass={cs.DC},
      url={https://arxiv.org/abs/2401.11181}, 
}

@misc{jain2025intelligentrouterllmworkloads,
      title={Intelligent Router for LLM Workloads: Improving Performance Through Workload-Aware Load Balancing}, 
      author={Kunal Jain and Anjaly Parayil and Ankur Mallick and Esha Choukse and Xiaoting Qin and Jue Zhang and Íñigo Goiri and Rujia Wang and Chetan Bansal and Victor Rühle and Anoop Kulkarni and Steve Kofsky and Saravan Rajmohan},
      year={2025},
      eprint={2408.13510},
      archivePrefix={arXiv},
      primaryClass={cs.DC},
      url={https://arxiv.org/abs/2408.13510}, 
}

@misc{geminiteam2025geminifamilyhighlycapable,
      title={Gemini: A Family of Highly Capable Multimodal Models}, 
      author={Gemini Team and Rohan Anil and Sebastian Borgeaud and Jean-Baptiste Alayrac and Jiahui Yu and Radu Soricut and Johan Schalkwyk and Andrew M. Dai and Anja Hauth and Katie Millican and David Silver and Melvin Johnson and Ioannis Antonoglou and Julian Schrittwieser and Amelia Glaese and Jilin Chen and Emily Pitler and Timothy Lillicrap and Angeliki Lazaridou and Orhan Firat and James Molloy and Michael Isard and Paul R. Barham and Tom Hennigan and Benjamin Lee and Fabio Viola and Malcolm Reynolds and Yuanzhong Xu and Ryan Doherty and Eli Collins and Clemens Meyer and Eliza Rutherford and Erica Moreira and Kareem Ayoub and Megha Goel and Jack Krawczyk and Cosmo Du and Ed Chi and Heng-Tze Cheng and Eric Ni and Purvi Shah and Patrick Kane and Betty Chan and Manaal Faruqui and Aliaksei Severyn and Hanzhao Lin and YaGuang Li and Yong Cheng and Abe Ittycheriah and Mahdis Mahdieh and Mia Chen and Pei Sun and Dustin Tran and Sumit Bagri and Balaji Lakshminarayanan and Jeremiah Liu and Andras Orban and Fabian Güra and Hao Zhou and Xinying Song and Aurelien Boffy and Harish Ganapathy and Steven Zheng and HyunJeong Choe and Ágoston Weisz and Tao Zhu and Yifeng Lu and Siddharth Gopal and Jarrod Kahn and Maciej Kula and Jeff Pitman and Rushin Shah and Emanuel Taropa and Majd Al Merey and Martin Baeuml and Zhifeng Chen and Laurent El Shafey and Yujing Zhang and Olcan Sercinoglu and George Tucker and Enrique Piqueras and Maxim Krikun and Iain Barr and Nikolay Savinov and Ivo Danihelka and Becca Roelofs and Anaïs White and Anders Andreassen and Tamara von Glehn and Lakshman Yagati and Mehran Kazemi and Lucas Gonzalez and Misha Khalman and Jakub Sygnowski and Alexandre Frechette and Charlotte Smith and Laura Culp and Lev Proleev and Yi Luan and Xi Chen and James Lottes and Nathan Schucher and Federico Lebron and Alban Rrustemi and Natalie Clay and Phil Crone and Tomas Kocisky and Jeffrey Zhao and Bartek Perz and Dian Yu and Heidi Howard and Adam Bloniarz and Jack W. Rae and Han Lu and Laurent Sifre and Marcello Maggioni and Fred Alcober and Dan Garrette and Megan Barnes and Shantanu Thakoor and Jacob Austin and Gabriel Barth-Maron and William Wong and Rishabh Joshi and Rahma Chaabouni and Deeni Fatiha and Arun Ahuja and Gaurav Singh Tomar and Evan Senter and Martin Chadwick and Ilya Kornakov and Nithya Attaluri and Iñaki Iturrate and Ruibo Liu and Yunxuan Li and Sarah Cogan and Jeremy Chen and Chao Jia and Chenjie Gu and Qiao Zhang and Jordan Grimstad and Ale Jakse Hartman and Xavier Garcia and Thanumalayan Sankaranarayana Pillai and Jacob Devlin and Michael Laskin and Diego de Las Casas and Dasha Valter and Connie Tao and Lorenzo Blanco and Adrià Puigdomènech Badia and David Reitter and Mianna Chen and Jenny Brennan and Clara Rivera and Sergey Brin and Shariq Iqbal and Gabriela Surita and Jane Labanowski and Abhi Rao and Stephanie Winkler and Emilio Parisotto and Yiming Gu and Kate Olszewska and Ravi Addanki and Antoine Miech and Annie Louis and Denis Teplyashin and Geoff Brown and Elliot Catt and Jan Balaguer and Jackie Xiang and Pidong Wang and Zoe Ashwood and Anton Briukhov and Albert Webson and Sanjay Ganapathy and Smit Sanghavi and Ajay Kannan and Ming-Wei Chang and Axel Stjerngren and Josip Djolonga and Yuting Sun and Ankur Bapna and Matthew Aitchison and Pedram Pejman and Henryk Michalewski and Tianhe Yu and Cindy Wang and Juliette Love and Junwhan Ahn and Dawn Bloxwich and Kehang Han and Peter Humphreys and Thibault Sellam and James Bradbury and Varun Godbole and Sina Samangooei and Bogdan Damoc and Alex Kaskasoli and Sébastien M. R. Arnold and Vijay Vasudevan and Shubham Agrawal and Jason Riesa and Dmitry Lepikhin and Richard Tanburn and Srivatsan Srinivasan and Hyeontaek Lim and Sarah Hodkinson and Pranav Shyam and Johan Ferret and Steven Hand and Ankush Garg and Tom Le Paine and Jian Li and Yujia Li and Minh Giang and Alexander Neitz and Zaheer Abbas and Sarah York and Machel Reid and Elizabeth Cole and Aakanksha Chowdhery and Dipanjan Das and Dominika Rogozińska and Vitaliy Nikolaev and Pablo Sprechmann and Zachary Nado and Lukas Zilka and Flavien Prost and Luheng He and Marianne Monteiro and Gaurav Mishra and Chris Welty and Josh Newlan and Dawei Jia and Miltiadis Allamanis and Clara Huiyi Hu and Raoul de Liedekerke and Justin Gilmer and Carl Saroufim and Shruti Rijhwani and Shaobo Hou and Disha Shrivastava and Anirudh Baddepudi and Alex Goldin and Adnan Ozturel and Albin Cassirer and Yunhan Xu and Daniel Sohn and Devendra Sachan and Reinald Kim Amplayo and Craig Swanson and Dessie Petrova and Shashi Narayan and Arthur Guez and Siddhartha Brahma and Jessica Landon and Miteyan Patel and Ruizhe Zhao and Kevin Villela and Luyu Wang and Wenhao Jia and Matthew Rahtz and Mai Giménez and Legg Yeung and James Keeling and Petko Georgiev and Diana Mincu and Boxi Wu and Salem Haykal and Rachel Saputro and Kiran Vodrahalli and James Qin and Zeynep Cankara and Abhanshu Sharma and Nick Fernando and Will Hawkins and Behnam Neyshabur and Solomon Kim and Adrian Hutter and Priyanka Agrawal and Alex Castro-Ros and George van den Driessche and Tao Wang and Fan Yang and Shuo-yiin Chang and Paul Komarek and Ross McIlroy and Mario Lučić and Guodong Zhang and Wael Farhan and Michael Sharman and Paul Natsev and Paul Michel and Yamini Bansal and Siyuan Qiao and Kris Cao and Siamak Shakeri and Christina Butterfield and Justin Chung and Paul Kishan Rubenstein and Shivani Agrawal and Arthur Mensch and Kedar Soparkar and Karel Lenc and Timothy Chung and Aedan Pope and Loren Maggiore and Jackie Kay and Priya Jhakra and Shibo Wang and Joshua Maynez and Mary Phuong and Taylor Tobin and Andrea Tacchetti and Maja Trebacz and Kevin Robinson and Yash Katariya and Sebastian Riedel and Paige Bailey and Kefan Xiao and Nimesh Ghelani and Lora Aroyo and Ambrose Slone and Neil Houlsby and Xuehan Xiong and Zhen Yang and Elena Gribovskaya and Jonas Adler and Mateo Wirth and Lisa Lee and Music Li and Thais Kagohara and Jay Pavagadhi and Sophie Bridgers and Anna Bortsova and Sanjay Ghemawat and Zafarali Ahmed and Tianqi Liu and Richard Powell and Vijay Bolina and Mariko Iinuma and Polina Zablotskaia and James Besley and Da-Woon Chung and Timothy Dozat and Ramona Comanescu and Xiance Si and Jeremy Greer and Guolong Su and Martin Polacek and Raphaël Lopez Kaufman and Simon Tokumine and Hexiang Hu and Elena Buchatskaya and Yingjie Miao and Mohamed Elhawaty and Aditya Siddhant and Nenad Tomasev and Jinwei Xing and Christina Greer and Helen Miller and Shereen Ashraf and Aurko Roy and Zizhao Zhang and Ada Ma and Angelos Filos and Milos Besta and Rory Blevins and Ted Klimenko and Chih-Kuan Yeh and Soravit Changpinyo and Jiaqi Mu and Oscar Chang and Mantas Pajarskas and Carrie Muir and Vered Cohen and Charline Le Lan and Krishna Haridasan and Amit Marathe and Steven Hansen and Sholto Douglas and Rajkumar Samuel and Mingqiu Wang and Sophia Austin and Chang Lan and Jiepu Jiang and Justin Chiu and Jaime Alonso Lorenzo and Lars Lowe Sjösund and Sébastien Cevey and Zach Gleicher and Thi Avrahami and Anudhyan Boral and Hansa Srinivasan and Vittorio Selo and Rhys May and Konstantinos Aisopos and Léonard Hussenot and Livio Baldini Soares and Kate Baumli and Michael B. Chang and Adrià Recasens and Ben Caine and Alexander Pritzel and Filip Pavetic and Fabio Pardo and Anita Gergely and Justin Frye and Vinay Ramasesh and Dan Horgan and Kartikeya Badola and Nora Kassner and Subhrajit Roy and Ethan Dyer and Víctor Campos Campos and Alex Tomala and Yunhao Tang and Dalia El Badawy and Elspeth White and Basil Mustafa and Oran Lang and Abhishek Jindal and Sharad Vikram and Zhitao Gong and Sergi Caelles and Ross Hemsley and Gregory Thornton and Fangxiaoyu Feng and Wojciech Stokowiec and Ce Zheng and Phoebe Thacker and Çağlar Ünlü and Zhishuai Zhang and Mohammad Saleh and James Svensson and Max Bileschi and Piyush Patil and Ankesh Anand and Roman Ring and Katerina Tsihlas and Arpi Vezer and Marco Selvi and Toby Shevlane and Mikel Rodriguez and Tom Kwiatkowski and Samira Daruki and Keran Rong and Allan Dafoe and Nicholas FitzGerald and Keren Gu-Lemberg and Mina Khan and Lisa Anne Hendricks and Marie Pellat and Vladimir Feinberg and James Cobon-Kerr and Tara Sainath and Maribeth Rauh and Sayed Hadi Hashemi and Richard Ives and Yana Hasson and Eric Noland and Yuan Cao and Nathan Byrd and Le Hou and Qingze Wang and Thibault Sottiaux and Michela Paganini and Jean-Baptiste Lespiau and Alexandre Moufarek and Samer Hassan and Kaushik Shivakumar and Joost van Amersfoort and Amol Mandhane and Pratik Joshi and Anirudh Goyal and Matthew Tung and Andrew Brock and Hannah Sheahan and Vedant Misra and Cheng Li and Nemanja Rakićević and Mostafa Dehghani and Fangyu Liu and Sid Mittal and Junhyuk Oh and Seb Noury and Eren Sezener and Fantine Huot and Matthew Lamm and Nicola De Cao and Charlie Chen and Sidharth Mudgal and Romina Stella and Kevin Brooks and Gautam Vasudevan and Chenxi Liu and Mainak Chain and Nivedita Melinkeri and Aaron Cohen and Venus Wang and Kristie Seymore and Sergey Zubkov and Rahul Goel and Summer Yue and Sai Krishnakumaran and Brian Albert and Nate Hurley and Motoki Sano and Anhad Mohananey and Jonah Joughin and Egor Filonov and Tomasz Kępa and Yomna Eldawy and Jiawern Lim and Rahul Rishi and Shirin Badiezadegan and Taylor Bos and Jerry Chang and Sanil Jain and Sri Gayatri Sundara Padmanabhan and Subha Puttagunta and Kalpesh Krishna and Leslie Baker and Norbert Kalb and Vamsi Bedapudi and Adam Kurzrok and Shuntong Lei and Anthony Yu and Oren Litvin and Xiang Zhou and Zhichun Wu and Sam Sobell and Andrea Siciliano and Alan Papir and Robby Neale and Jonas Bragagnolo and Tej Toor and Tina Chen and Valentin Anklin and Feiran Wang and Richie Feng and Milad Gholami and Kevin Ling and Lijuan Liu and Jules Walter and Hamid Moghaddam and Arun Kishore and Jakub Adamek and Tyler Mercado and Jonathan Mallinson and Siddhinita Wandekar and Stephen Cagle and Eran Ofek and Guillermo Garrido and Clemens Lombriser and Maksim Mukha and Botu Sun and Hafeezul Rahman Mohammad and Josip Matak and Yadi Qian and Vikas Peswani and Pawel Janus and Quan Yuan and Leif Schelin and Oana David and Ankur Garg and Yifan He and Oleksii Duzhyi and Anton Älgmyr and Timothée Lottaz and Qi Li and Vikas Yadav and Luyao Xu and Alex Chinien and Rakesh Shivanna and Aleksandr Chuklin and Josie Li and Carrie Spadine and Travis Wolfe and Kareem Mohamed and Subhabrata Das and Zihang Dai and Kyle He and Daniel von Dincklage and Shyam Upadhyay and Akanksha Maurya and Luyan Chi and Sebastian Krause and Khalid Salama and Pam G Rabinovitch and Pavan Kumar Reddy M and Aarush Selvan and Mikhail Dektiarev and Golnaz Ghiasi and Erdem Guven and Himanshu Gupta and Boyi Liu and Deepak Sharma and Idan Heimlich Shtacher and Shachi Paul and Oscar Akerlund and François-Xavier Aubet and Terry Huang and Chen Zhu and Eric Zhu and Elico Teixeira and Matthew Fritze and Francesco Bertolini and Liana-Eleonora Marinescu and Martin Bölle and Dominik Paulus and Khyatti Gupta and Tejasi Latkar and Max Chang and Jason Sanders and Roopa Wilson and Xuewei Wu and Yi-Xuan Tan and Lam Nguyen Thiet and Tulsee Doshi and Sid Lall and Swaroop Mishra and Wanming Chen and Thang Luong and Seth Benjamin and Jasmine Lee and Ewa Andrejczuk and Dominik Rabiej and Vipul Ranjan and Krzysztof Styrc and Pengcheng Yin and Jon Simon and Malcolm Rose Harriott and Mudit Bansal and Alexei Robsky and Geoff Bacon and David Greene and Daniil Mirylenka and Chen Zhou and Obaid Sarvana and Abhimanyu Goyal and Samuel Andermatt and Patrick Siegler and Ben Horn and Assaf Israel and Francesco Pongetti and Chih-Wei "Louis" Chen and Marco Selvatici and Pedro Silva and Kathie Wang and Jackson Tolins and Kelvin Guu and Roey Yogev and Xiaochen Cai and Alessandro Agostini and Maulik Shah and Hung Nguyen and Noah O Donnaile and Sébastien Pereira and Linda Friso and Adam Stambler and Adam Kurzrok and Chenkai Kuang and Yan Romanikhin and Mark Geller and ZJ Yan and Kane Jang and Cheng-Chun Lee and Wojciech Fica and Eric Malmi and Qijun Tan and Dan Banica and Daniel Balle and Ryan Pham and Yanping Huang and Diana Avram and Hongzhi Shi and Jasjot Singh and Chris Hidey and Niharika Ahuja and Pranab Saxena and Dan Dooley and Srividya Pranavi Potharaju and Eileen O'Neill and Anand Gokulchandran and Ryan Foley and Kai Zhao and Mike Dusenberry and Yuan Liu and Pulkit Mehta and Ragha Kotikalapudi and Chalence Safranek-Shrader and Andrew Goodman and Joshua Kessinger and Eran Globen and Prateek Kolhar and Chris Gorgolewski and Ali Ibrahim and Yang Song and Ali Eichenbaum and Thomas Brovelli and Sahitya Potluri and Preethi Lahoti and Cip Baetu and Ali Ghorbani and Charles Chen and Andy Crawford and Shalini Pal and Mukund Sridhar and Petru Gurita and Asier Mujika and Igor Petrovski and Pierre-Louis Cedoz and Chenmei Li and Shiyuan Chen and Niccolò Dal Santo and Siddharth Goyal and Jitesh Punjabi and Karthik Kappaganthu and Chester Kwak and Pallavi LV and Sarmishta Velury and Himadri Choudhury and Jamie Hall and Premal Shah and Ricardo Figueira and Matt Thomas and Minjie Lu and Ting Zhou and Chintu Kumar and Thomas Jurdi and Sharat Chikkerur and Yenai Ma and Adams Yu and Soo Kwak and Victor Ähdel and Sujeevan Rajayogam and Travis Choma and Fei Liu and Aditya Barua and Colin Ji and Ji Ho Park and Vincent Hellendoorn and Alex Bailey and Taylan Bilal and Huanjie Zhou and Mehrdad Khatir and Charles Sutton and Wojciech Rzadkowski and Fiona Macintosh and Roopali Vij and Konstantin Shagin and Paul Medina and Chen Liang and Jinjing Zhou and Pararth Shah and Yingying Bi and Attila Dankovics and Shipra Banga and Sabine Lehmann and Marissa Bredesen and Zifan Lin and John Eric Hoffmann and Jonathan Lai and Raynald Chung and Kai Yang and Nihal Balani and Arthur Bražinskas and Andrei Sozanschi and Matthew Hayes and Héctor Fernández Alcalde and Peter Makarov and Will Chen and Antonio Stella and Liselotte Snijders and Michael Mandl and Ante Kärrman and Paweł Nowak and Xinyi Wu and Alex Dyck and Krishnan Vaidyanathan and Raghavender R and Jessica Mallet and Mitch Rudominer and Eric Johnston and Sushil Mittal and Akhil Udathu and Janara Christensen and Vishal Verma and Zach Irving and Andreas Santucci and Gamaleldin Elsayed and Elnaz Davoodi and Marin Georgiev and Ian Tenney and Nan Hua and Geoffrey Cideron and Edouard Leurent and Mahmoud Alnahlawi and Ionut Georgescu and Nan Wei and Ivy Zheng and Dylan Scandinaro and Heinrich Jiang and Jasper Snoek and Mukund Sundararajan and Xuezhi Wang and Zack Ontiveros and Itay Karo and Jeremy Cole and Vinu Rajashekhar and Lara Tumeh and Eyal Ben-David and Rishub Jain and Jonathan Uesato and Romina Datta and Oskar Bunyan and Shimu Wu and John Zhang and Piotr Stanczyk and Ye Zhang and David Steiner and Subhajit Naskar and Michael Azzam and Matthew Johnson and Adam Paszke and Chung-Cheng Chiu and Jaume Sanchez Elias and Afroz Mohiuddin and Faizan Muhammad and Jin Miao and Andrew Lee and Nino Vieillard and Jane Park and Jiageng Zhang and Jeff Stanway and Drew Garmon and Abhijit Karmarkar and Zhe Dong and Jong Lee and Aviral Kumar and Luowei Zhou and Jonathan Evens and William Isaac and Geoffrey Irving and Edward Loper and Michael Fink and Isha Arkatkar and Nanxin Chen and Izhak Shafran and Ivan Petrychenko and Zhe Chen and Johnson Jia and Anselm Levskaya and Zhenkai Zhu and Peter Grabowski and Yu Mao and Alberto Magni and Kaisheng Yao and Javier Snaider and Norman Casagrande and Evan Palmer and Paul Suganthan and Alfonso Castaño and Irene Giannoumis and Wooyeol Kim and Mikołaj Rybiński and Ashwin Sreevatsa and Jennifer Prendki and David Soergel and Adrian Goedeckemeyer and Willi Gierke and Mohsen Jafari and Meenu Gaba and Jeremy Wiesner and Diana Gage Wright and Yawen Wei and Harsha Vashisht and Yana Kulizhskaya and Jay Hoover and Maigo Le and Lu Li and Chimezie Iwuanyanwu and Lu Liu and Kevin Ramirez and Andrey Khorlin and Albert Cui and Tian LIN and Marcus Wu and Ricardo Aguilar and Keith Pallo and Abhishek Chakladar and Ginger Perng and Elena Allica Abellan and Mingyang Zhang and Ishita Dasgupta and Nate Kushman and Ivo Penchev and Alena Repina and Xihui Wu and Tom van der Weide and Priya Ponnapalli and Caroline Kaplan and Jiri Simsa and Shuangfeng Li and Olivier Dousse and Fan Yang and Jeff Piper and Nathan Ie and Rama Pasumarthi and Nathan Lintz and Anitha Vijayakumar and Daniel Andor and Pedro Valenzuela and Minnie Lui and Cosmin Paduraru and Daiyi Peng and Katherine Lee and Shuyuan Zhang and Somer Greene and Duc Dung Nguyen and Paula Kurylowicz and Cassidy Hardin and Lucas Dixon and Lili Janzer and Kiam Choo and Ziqiang Feng and Biao Zhang and Achintya Singhal and Dayou Du and Dan McKinnon and Natasha Antropova and Tolga Bolukbasi and Orgad Keller and David Reid and Daniel Finchelstein and Maria Abi Raad and Remi Crocker and Peter Hawkins and Robert Dadashi and Colin Gaffney and Ken Franko and Anna Bulanova and Rémi Leblond and Shirley Chung and Harry Askham and Luis C. Cobo and Kelvin Xu and Felix Fischer and Jun Xu and Christina Sorokin and Chris Alberti and Chu-Cheng Lin and Colin Evans and Alek Dimitriev and Hannah Forbes and Dylan Banarse and Zora Tung and Mark Omernick and Colton Bishop and Rachel Sterneck and Rohan Jain and Jiawei Xia and Ehsan Amid and Francesco Piccinno and Xingyu Wang and Praseem Banzal and Daniel J. Mankowitz and Alex Polozov and Victoria Krakovna and Sasha Brown and MohammadHossein Bateni and Dennis Duan and Vlad Firoiu and Meghana Thotakuri and Tom Natan and Matthieu Geist and Ser tan Girgin and Hui Li and Jiayu Ye and Ofir Roval and Reiko Tojo and Michael Kwong and James Lee-Thorp and Christopher Yew and Danila Sinopalnikov and Sabela Ramos and John Mellor and Abhishek Sharma and Kathy Wu and David Miller and Nicolas Sonnerat and Denis Vnukov and Rory Greig and Jennifer Beattie and Emily Caveness and Libin Bai and Julian Eisenschlos and Alex Korchemniy and Tomy Tsai and Mimi Jasarevic and Weize Kong and Phuong Dao and Zeyu Zheng and Frederick Liu and Fan Yang and Rui Zhu and Tian Huey Teh and Jason Sanmiya and Evgeny Gladchenko and Nejc Trdin and Daniel Toyama and Evan Rosen and Sasan Tavakkol and Linting Xue and Chen Elkind and Oliver Woodman and John Carpenter and George Papamakarios and Rupert Kemp and Sushant Kafle and Tanya Grunina and Rishika Sinha and Alice Talbert and Diane Wu and Denese Owusu-Afriyie and Cosmo Du and Chloe Thornton and Jordi Pont-Tuset and Pradyumna Narayana and Jing Li and Saaber Fatehi and John Wieting and Omar Ajmeri and Benigno Uria and Yeongil Ko and Laura Knight and Amélie Héliou and Ning Niu and Shane Gu and Chenxi Pang and Yeqing Li and Nir Levine and Ariel Stolovich and Rebeca Santamaria-Fernandez and Sonam Goenka and Wenny Yustalim and Robin Strudel and Ali Elqursh and Charlie Deck and Hyo Lee and Zonglin Li and Kyle Levin and Raphael Hoffmann and Dan Holtmann-Rice and Olivier Bachem and Sho Arora and Christy Koh and Soheil Hassas Yeganeh and Siim Põder and Mukarram Tariq and Yanhua Sun and Lucian Ionita and Mojtaba Seyedhosseini and Pouya Tafti and Zhiyu Liu and Anmol Gulati and Jasmine Liu and Xinyu Ye and Bart Chrzaszcz and Lily Wang and Nikhil Sethi and Tianrun Li and Ben Brown and Shreya Singh and Wei Fan and Aaron Parisi and Joe Stanton and Vinod Koverkathu and Christopher A. Choquette-Choo and Yunjie Li and TJ Lu and Abe Ittycheriah and Prakash Shroff and Mani Varadarajan and Sanaz Bahargam and Rob Willoughby and David Gaddy and Guillaume Desjardins and Marco Cornero and Brona Robenek and Bhavishya Mittal and Ben Albrecht and Ashish Shenoy and Fedor Moiseev and Henrik Jacobsson and Alireza Ghaffarkhah and Morgane Rivière and Alanna Walton and Clément Crepy and Alicia Parrish and Zongwei Zhou and Clement Farabet and Carey Radebaugh and Praveen Srinivasan and Claudia van der Salm and Andreas Fidjeland and Salvatore Scellato and Eri Latorre-Chimoto and Hanna Klimczak-Plucińska and David Bridson and Dario de Cesare and Tom Hudson and Piermaria Mendolicchio and Lexi Walker and Alex Morris and Matthew Mauger and Alexey Guseynov and Alison Reid and Seth Odoom and Lucia Loher and Victor Cotruta and Madhavi Yenugula and Dominik Grewe and Anastasia Petrushkina and Tom Duerig and Antonio Sanchez and Steve Yadlowsky and Amy Shen and Amir Globerson and Lynette Webb and Sahil Dua and Dong Li and Surya Bhupatiraju and Dan Hurt and Haroon Qureshi and Ananth Agarwal and Tomer Shani and Matan Eyal and Anuj Khare and Shreyas Rammohan Belle and Lei Wang and Chetan Tekur and Mihir Sanjay Kale and Jinliang Wei and Ruoxin Sang and Brennan Saeta and Tyler Liechty and Yi Sun and Yao Zhao and Stephan Lee and Pandu Nayak and Doug Fritz and Manish Reddy Vuyyuru and John Aslanides and Nidhi Vyas and Martin Wicke and Xiao Ma and Evgenii Eltyshev and Nina Martin and Hardie Cate and James Manyika and Keyvan Amiri and Yelin Kim and Xi Xiong and Kai Kang and Florian Luisier and Nilesh Tripuraneni and David Madras and Mandy Guo and Austin Waters and Oliver Wang and Joshua Ainslie and Jason Baldridge and Han Zhang and Garima Pruthi and Jakob Bauer and Feng Yang and Riham Mansour and Jason Gelman and Yang Xu and George Polovets and Ji Liu and Honglong Cai and Warren Chen and XiangHai Sheng and Emily Xue and Sherjil Ozair and Christof Angermueller and Xiaowei Li and Anoop Sinha and Weiren Wang and Julia Wiesinger and Emmanouil Koukoumidis and Yuan Tian and Anand Iyer and Madhu Gurumurthy and Mark Goldenson and Parashar Shah and MK Blake and Hongkun Yu and Anthony Urbanowicz and Jennimaria Palomaki and Chrisantha Fernando and Ken Durden and Harsh Mehta and Nikola Momchev and Elahe Rahimtoroghi and Maria Georgaki and Amit Raul and Sebastian Ruder and Morgan Redshaw and Jinhyuk Lee and Denny Zhou and Komal Jalan and Dinghua Li and Blake Hechtman and Parker Schuh and Milad Nasr and Kieran Milan and Vladimir Mikulik and Juliana Franco and Tim Green and Nam Nguyen and Joe Kelley and Aroma Mahendru and Andrea Hu and Joshua Howland and Ben Vargas and Jeffrey Hui and Kshitij Bansal and Vikram Rao and Rakesh Ghiya and Emma Wang and Ke Ye and Jean Michel Sarr and Melanie Moranski Preston and Madeleine Elish and Steve Li and Aakash Kaku and Jigar Gupta and Ice Pasupat and Da-Cheng Juan and Milan Someswar and Tejvi M. and Xinyun Chen and Aida Amini and Alex Fabrikant and Eric Chu and Xuanyi Dong and Amruta Muthal and Senaka Buthpitiya and Sarthak Jauhari and Nan Hua and Urvashi Khandelwal and Ayal Hitron and Jie Ren and Larissa Rinaldi and Shahar Drath and Avigail Dabush and Nan-Jiang Jiang and Harshal Godhia and Uli Sachs and Anthony Chen and Yicheng Fan and Hagai Taitelbaum and Hila Noga and Zhuyun Dai and James Wang and Chen Liang and Jenny Hamer and Chun-Sung Ferng and Chenel Elkind and Aviel Atias and Paulina Lee and Vít Listík and Mathias Carlen and Jan van de Kerkhof and Marcin Pikus and Krunoslav Zaher and Paul Müller and Sasha Zykova and Richard Stefanec and Vitaly Gatsko and Christoph Hirnschall and Ashwin Sethi and Xingyu Federico Xu and Chetan Ahuja and Beth Tsai and Anca Stefanoiu and Bo Feng and Keshav Dhandhania and Manish Katyal and Akshay Gupta and Atharva Parulekar and Divya Pitta and Jing Zhao and Vivaan Bhatia and Yashodha Bhavnani and Omar Alhadlaq and Xiaolin Li and Peter Danenberg and Dennis Tu and Alex Pine and Vera Filippova and Abhipso Ghosh and Ben Limonchik and Bhargava Urala and Chaitanya Krishna Lanka and Derik Clive and Yi Sun and Edward Li and Hao Wu and Kevin Hongtongsak and Ianna Li and Kalind Thakkar and Kuanysh Omarov and Kushal Majmundar and Michael Alverson and Michael Kucharski and Mohak Patel and Mudit Jain and Maksim Zabelin and Paolo Pelagatti and Rohan Kohli and Saurabh Kumar and Joseph Kim and Swetha Sankar and Vineet Shah and Lakshmi Ramachandruni and Xiangkai Zeng and Ben Bariach and Laura Weidinger and Tu Vu and Alek Andreev and Antoine He and Kevin Hui and Sheleem Kashem and Amar Subramanya and Sissie Hsiao and Demis Hassabis and Koray Kavukcuoglu and Adam Sadovsky and Quoc Le and Trevor Strohman and Yonghui Wu and Slav Petrov and Jeffrey Dean and Oriol Vinyals},
      year={2025},
      eprint={2312.11805},
      archivePrefix={arXiv},
      primaryClass={cs.CL},
      url={https://arxiv.org/abs/2312.11805}, 
}
\bibliographystyle{icml2026}

\clearpage
\appendix

\section{Configuration Space}
\label{app:configs}

\begin{table*}[h]
\caption{Complete list of 17 configurations evaluated in our experiments. The routing parameter $x$ determines the fraction of AP operations routed to D nodes.}
\label{tab:all-configs}
\centering
\small
\begin{tabular}{llcc}
\toprule
\textbf{Category} & \textbf{Configurations} & \textbf{$x$} & \textbf{Count} \\
\midrule
Replica & 4R & N/A & 1 \\
\midrule
$x{=}0$ (PD) & 1P\_3D, 2P\_2D, 3P\_1D & 0 & 3 \\
\midrule
$x{=}1$ (Full AP-to-D) & 1P\_3D, 2P\_2D, 3P\_1D & 1 & 3 \\
\midrule
$0{<}x{<}1$ (partial AP routing) & 1P\_3D, 2P\_2D & $\frac{1}{3}$, $\frac{2}{3}$, $\frac{1}{2}$ & 3 \\
\midrule
Hybrid & 1R\_1P\_2D, 1R\_2P\_1D, 2R\_1P\_1D, etc. & 0, 1, $\frac{1}{2}$ & 7 \\
\bottomrule
\end{tabular}
\end{table*}

All configurations use a fixed budget of 4 GPUs. Notation: xP = x Prefill-only nodes, xD = x Decode nodes, xR = x Replica nodes. The routing parameter $x \in [0,1]$ determines D's Turn 2+ behavior.

\paragraph{Why Hybrid Configurations Are Excluded from Main Analysis.}
Our experiments include 7 hybrid configurations that combine R (Replica) nodes with P/D nodes, as listed in \Cref{tab:all-configs}.
However, we exclude hybrid configurations from our main analysis for two reasons.

First, R nodes do not benefit from disaggregation's workload isolation.
Unlike dedicated P and D nodes, an R node handles both prefill and decode operations on the same GPU.
When R processes a prefill operation, it interferes with its own ongoing decode batches, the very problem that PD disaggregation was designed to solve.
This means the routing parameter $x$ (which controls AP routing to D nodes) does not apply to R nodes, making hybrid configurations orthogonal to our focus on AP routing optimization.

Second, hybrid configurations reduce the pool of dedicated P/D resources.
For example, 1R\_1P\_2D allocates only 1 GPU to prefill (vs.\ 2 in 2P\_2D), creating potential prefill bottlenecks under high load.
In our experiments, hybrid configurations rarely achieved the best performance on any metric, winning only 6.1\% of TTFT, 12.2\% of TPOT, and 16.1\% of throughput test points.

\section{Implementation Details}
\label{app:implementation}

This appendix provides implementation details for the PPD routing system described in \Cref{sec:ppd-design}.

\subsection{Decision-Engine Discretization}
\label{app:discretization}

The PPD lookup table indexes Phase-1 measurements along three discrete axes:
\begin{itemize}[leftmargin=*,nosep]
    \item \emph{Context class} bins the accumulated context length $n_{\text{ctx}}$ (cumulative tokens from prior turns) into \texttt{small} ($\leq$512), \texttt{large} (512--4096), and \texttt{huge} ($>$4096) tokens.
    \item \emph{Workload type} classifies the input/output token ratio $n_{\text{in}}/n_{\text{out}}$ into nine categories spanning decode-heavy, balanced, and prefill-heavy regimes (matching the synthetic grid in \Cref{subsec:config-space}).
    \item \emph{QPS bin} snaps the current system QPS to the nearest benchmark level used in Phase~1.
\end{itemize}
At inference time, an incoming request with feature tuple $\psi = (t, n_{\text{in}}, n_{\text{out}}, n_{\text{ctx}}, q)$ (turn number, new-input tokens, expected output tokens, accumulated context, current QPS) is mapped to the nearest grid cell $\hat{\psi}$ along these axes, and the precomputed $x^*(\hat{\psi})$ is returned.

\subsection{KV Transfer Protocol}

Our implementation reuses vLLM's disaggregated serving infrastructure.
P nodes run with \texttt{kv\_role=kv\_producer}, generating KV caches and sending them via ZeroMQ.
Decode servers run with \texttt{kv\_role=kv\_consumer}, receiving KV caches and storing them in local prefix cache for Turn 2+ processing when $x{>}0$.
The transfer protocol follows vLLM's standard disaggregated serving format, requiring no custom modifications.

\subsection{Session Management}

The routing proxy maintains a session table as an in-memory dictionary:
\begin{verbatim}
session_table[conv_hash] = {
    "turn_count": int,
    "assigned_pd": str,
    "last_access": timestamp
}
\end{verbatim}
Conversation identifiers are computed as MD5 hashes of the first user message.
Sessions are evicted after 60 minutes of inactivity, aligning with vLLM's default prefix cache TTL.

\subsection{Service Discovery}

Backend servers register with the proxy via ZeroMQ heartbeats every 10 seconds.
The proxy maintains instance lists for each server type (P, D, R) and removes servers after 30 seconds without a heartbeat.
This enables dynamic scaling and automatic failure recovery.

\subsection{Proxy Implementation}

The routing proxy is implemented using Quart (async Flask) and handles:
\begin{itemize}
    \item Request parsing: Extract conversation context and determine turn number
    \item Session lookup/creation: Manage conversation-to-server mappings
    \item Request forwarding: Stream responses from backend to client
    \item Statistics collection: Track routing decisions and latency metrics
\end{itemize}

\subsection{Bandwidth Simulation}
\label{app:bandwidth-sim}

To emulate slower interconnects without multi-node hardware (\Cref{subsec:scaling-sim}), we inject a calibrated extra delay on the decode-side receive path of vLLM's P2P NCCL connector. For each PD-routed request, the injected delay is
\begin{equation}
    \label{eq:simulated-delay}
    \Delta t \;=\; \max\!\Big(0,\;\tfrac{B(\psi)}{\beta_{\text{target}}} - t_{\text{NVLink}}\Big),
\end{equation}
where $B(\psi) = n_{\text{tokens}} \cdot s_{\text{kv}}$ is the request's KV cache footprint, $\beta_{\text{target}}$ is the simulated bandwidth, and $t_{\text{NVLink}}$ is the actual NVLink transfer time (subtracted to avoid double-counting). For Llama-3.1-8B (BF16, GQA with 8 KV heads, 32 layers) the per-token footprint is $s_{\text{kv}} = 128$~KiB; at the WildChat P90 context length of 5{,}115 tokens, each Turn 2+ transfer in the $x{=}0$ path moves $\sim$670~MB, about $\sim$4.5~ms over NVLink, $\sim$27~ms over InfiniBand HDR (25~GB/s), and $\sim$67~ms over 100GbE. PPD's local append-prefill path is unaffected by the injection.

\section{Additional Experimental Results}
\label{app:additional-results}

\subsection{Extended Interference Analysis}

The main paper (\Cref{fig:interference-tpot}) demonstrates the interference gap between full prefill and append-prefill with a single concurrent prefill operation. Here we extend this analysis to examine robustness under higher concurrency and longer context lengths.

\begin{figure}[h]
    \centering
    \includegraphics[width=0.85\columnwidth]{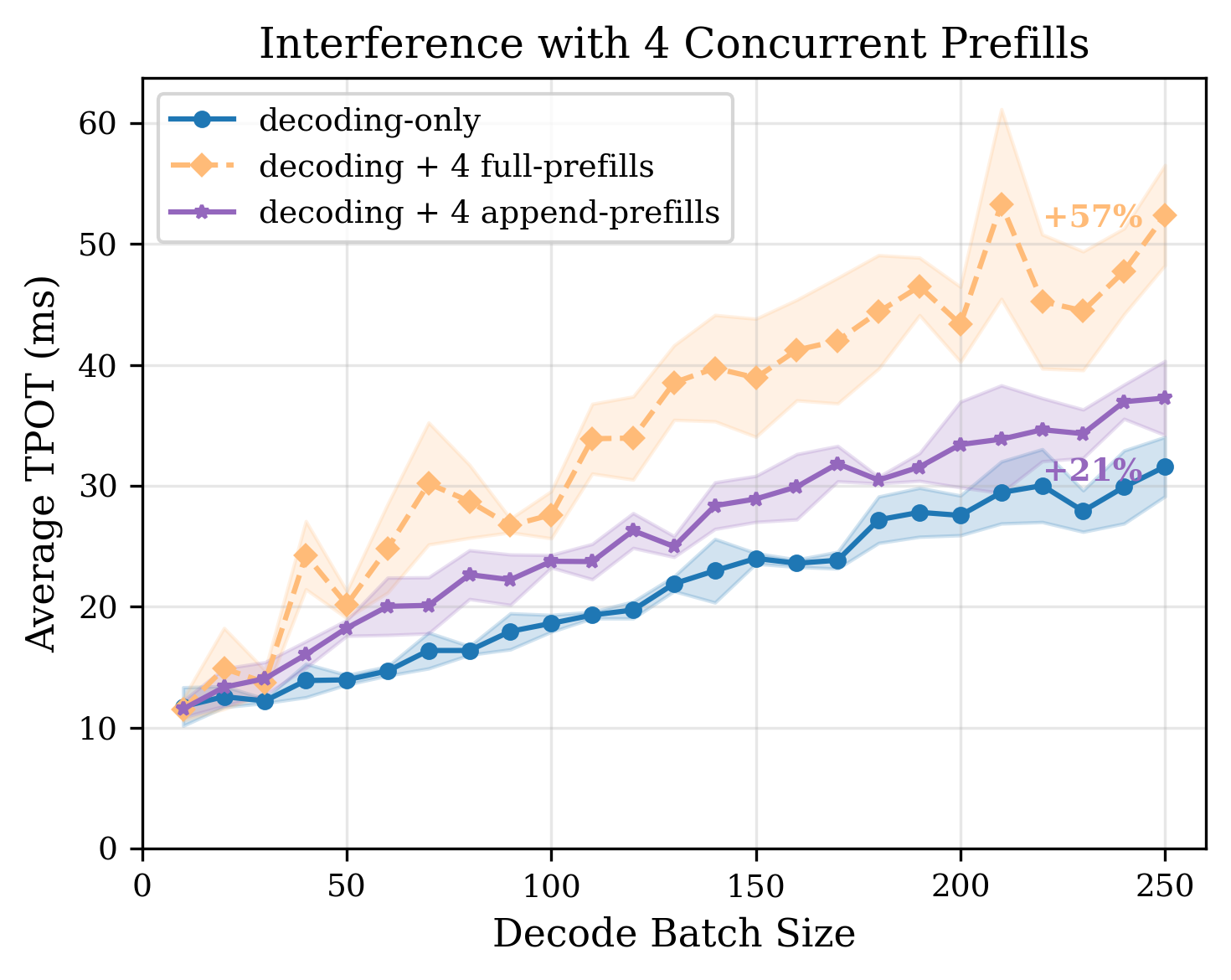}
    \caption{
        \textbf{Prefill-decode interference with 4 concurrent prefills.}
        Same setup as \Cref{fig:interference-tpot} but with 4 concurrent prefill operations.
        Full prefill causes $\sim$57\% slowdown at batch size 200; append-prefill remains within $\sim$21\% of baseline.
    }
    \label{fig:interference-4prefills}
\end{figure}

\Cref{fig:interference-4prefills} increases concurrency to 4 simultaneous prefill operations. The interference gap persists: full prefill causes $\sim$57\% TPOT degradation at batch size 200, while append-prefill remains within $\sim$21\% of baseline. This confirms that the fundamental difference in interference characteristics is not an artifact of low concurrency.

\begin{figure}[h]
    \centering
    \includegraphics[width=0.85\columnwidth]{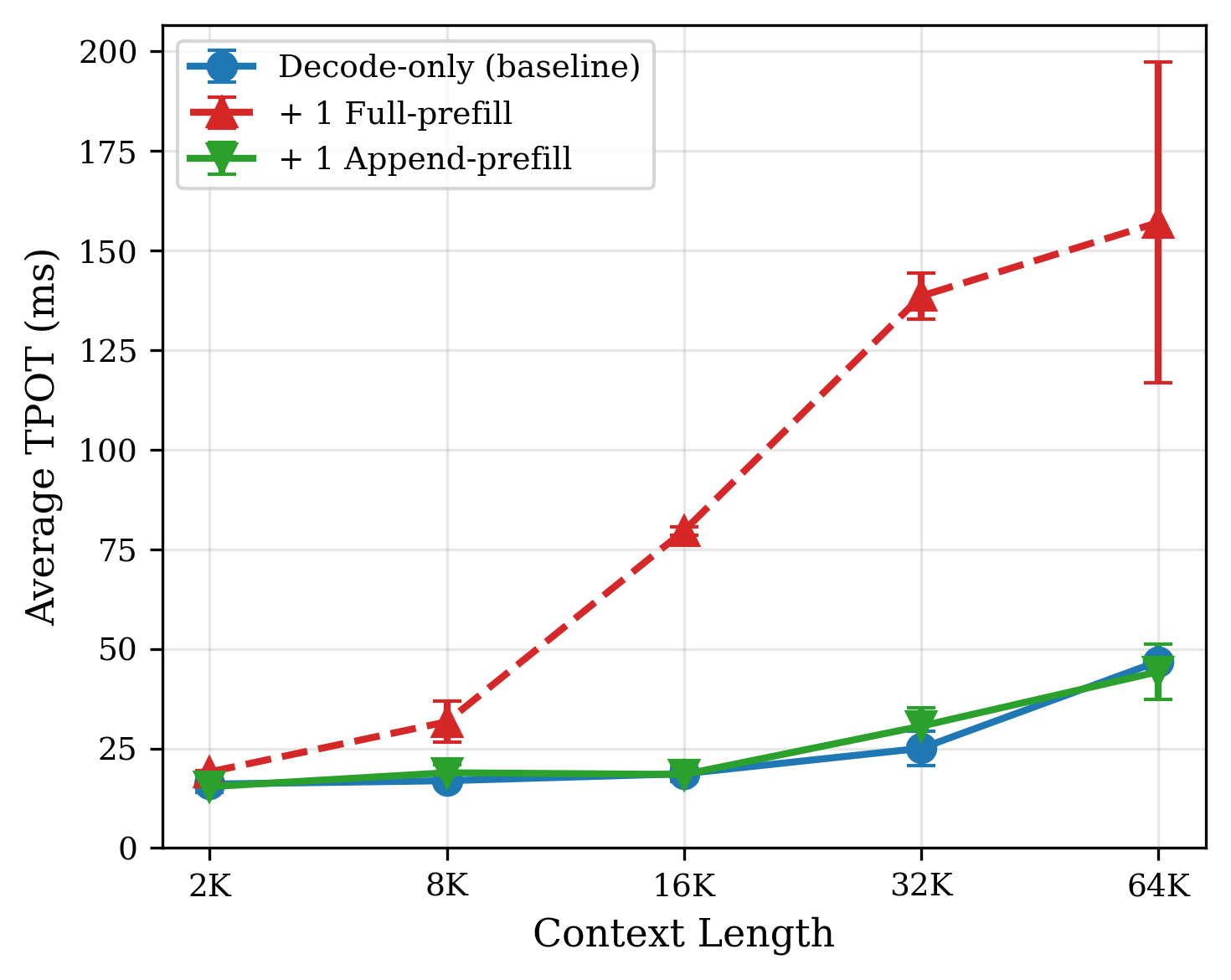}
    \caption{
        \textbf{Interference scaling with context length.}
        Full prefill interference grows to 3--4$\times$ at 32K tokens; append-prefill stays below 25\% even at 64K.
    }
    \label{fig:interference-sensitivity}
\end{figure}

\Cref{fig:interference-sensitivity} varies context length from 2K to 64K tokens. Full prefill interference grows quadratically, reaching 3--4$\times$ slowdown at 32K tokens. In contrast, append-prefill interference remains below 25\% even at 64K tokens, validating that append-prefill scales linearly with context length.

\subsection{Pareto Analysis}

\begin{figure*}[h]
    \centering
    \includegraphics[width=0.8\textwidth]{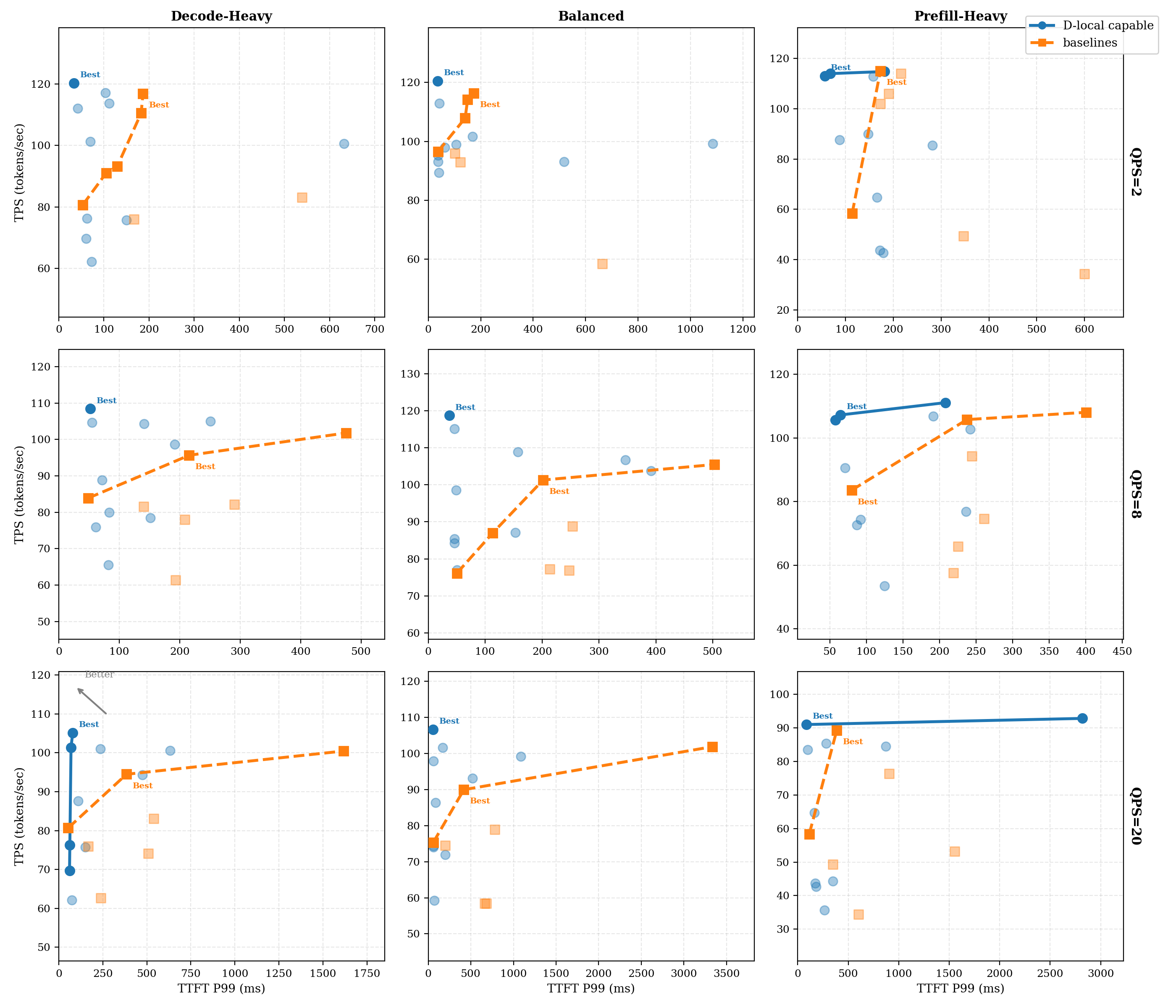}
    \caption{
        \textbf{P99 TTFT vs.\ TPS Pareto frontiers.}
        Columns: Decode-heavy, Balanced, Prefill-heavy workloads.
        Rows: QPS=2, 8, 20.
        D-local capable configurations (blue) achieve lower TTFT while maintaining competitive TPS, with the advantage most pronounced under high QPS and decode-heavy workloads.
    }
    \label{fig:pareto-small-context}
\end{figure*}

\Cref{fig:pareto-small-context} presents the complete Pareto frontier analysis across all workload types and QPS levels evaluated in our benchmark. Each subplot shows P99 TTFT versus throughput (TPS) for the 10 core configurations; the visualization complements the aggregate winner-distribution in \Cref{tab:winner-distribution} by exposing panel-level structure.

Two structural patterns deserve attention. Reading across columns (decode-heavy, balanced, prefill-heavy), the absolute TTFT scale grows because prefill-heavy mixes amplify both the KV transfer cost on the $x{=}0$ side and the append-prefill cost on the $x{=}1$ side, widening the spread of frontier points by roughly an order of magnitude between leftmost and rightmost columns at fixed QPS. Reading down rows (QPS=2, 8, 20), the orange baseline points fan out toward higher TTFT as transfer queuing accumulates, while the blue D-local-capable points stay clustered closer to the lower-TTFT region, particularly under decode-heavy mixes where the local prefix cache is hottest and append-prefill is cheapest.

The 9-panel grid refines the no-universal-best finding (\Cref{subsec:no-universal-best}): even within the two-axis TTFT vs.\ TPS view, no single configuration occupies every Pareto frontier. The optimum slides toward $x{=}1$ as QPS rises and the workload becomes more decode-heavy, while balanced and prefill-heavy panels at low QPS leave room for $x{=}0$ to compete on TPS. PPD's per-request choice rule (\Cref{eq:score}) is the natural way to reconcile this: it picks the locally favorable point on each panel rather than committing to one globally.

\subsection{Scaling Validation}

\begin{figure}[h]
    \centering
    \includegraphics[width=\columnwidth]{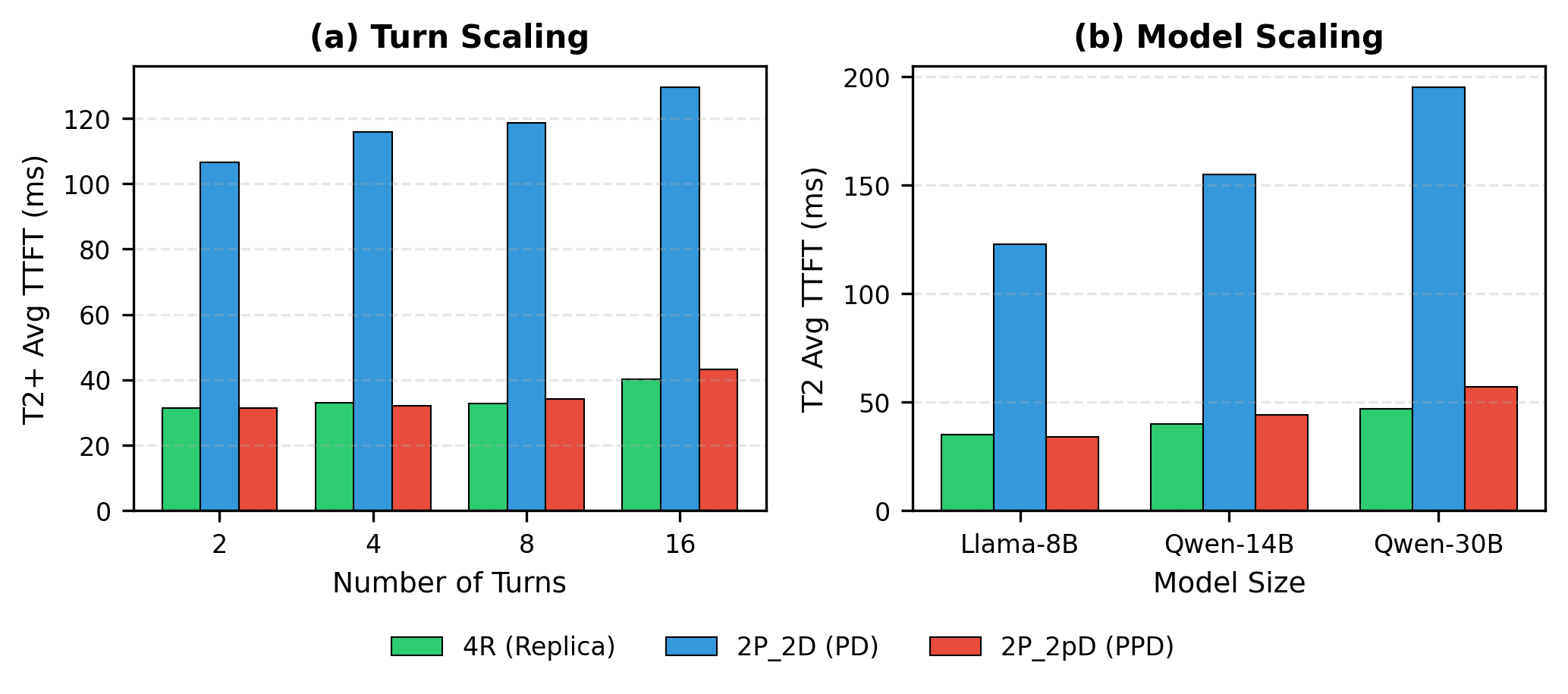}
    \caption{
        \textbf{$x{=}1$ maintains advantage across turns and model sizes.}
        (a) Turn scaling: Turn 2+ TTFT across 2--16 turns.
        (b) Model scaling: Turn 2 TTFT across 8B--30B models.
        $x{=}1$ closely matches Replica performance while preserving disaggregation benefits; $x{=}0$ degrades sharply with context growth.
    }
    \label{fig:validation}
\end{figure}

A natural question is whether the $x{=}1$ advantage observed in our 8B experiments generalizes to longer conversations and larger models. \Cref{fig:validation} addresses this by measuring Turn 2+ TTFT across (a) 2--16 turn conversations and (b) 8B--30B model sizes.

The results confirm that $x{=}1$ maintains its advantage across both dimensions. For turn scaling, the TTFT gap between $x{=}0$ and $x{=}1$ widens as conversations grow longer, because accumulated context amplifies KV transfer overhead. For model scaling, the relative improvement remains stable at $\sim$70\%, indicating that the benefit stems from architectural properties rather than model-specific characteristics.

Our comprehensive experiments focus on 8B models for practical reasons: replica configurations require the full model to fit on a single GPU, and each configuration switch requires a complete server restart. Testing 17 configurations across 180 workload-QPS combinations with larger models would incur prohibitive initialization overhead.

\subsection{Failure Rate Analysis}

Beyond latency metrics, system reliability is critical for production deployments. \Cref{tab:failure-rates} reports request failure rates (timeout $>$ 30s) across configurations at high QPS levels.

Two patterns emerge. First, configurations with extreme P:D ratios (3P\_1D) are inherently fragile: with only one decode GPU, any queuing at the decode stage cascades into timeouts. Second, within each P:D ratio, $x{=}1$ consistently achieves lower failure rates than $x{=}0$. For example, 2P\_2D with $x{=}1$ maintains 0\% failure rate at QPS=12 where $x{=}0$ already shows 6\% failures. This reliability advantage stems from reduced network contention: $x{=}1$ eliminates Turn 2+ KV transfers, freeing bandwidth for Turn 1 transfers that cannot be avoided.

\begin{table}[h]
\caption{Failure rates at high QPS. $x{=}1$ configurations consistently show lower failure rates than their $x{=}0$ counterparts.}
\label{tab:failure-rates}
\centering
\small
\begin{tabular}{lcccc}
\toprule
\textbf{Config} & \textbf{QPS=8} & \textbf{QPS=12} & \textbf{QPS=16} & \textbf{QPS=20} \\
\midrule
3P\_1D ($x{=}0$) & 11\% & 44\% & 61\% & 89\% \\
3P\_1D ($x{=}1$) & 6\% & 22\% & 44\% & 67\% \\
\midrule
2P\_2D ($x{=}0$) & 0\% & 6\% & 11\% & 22\% \\
2P\_2D ($x{=}1$) & 0\% & \textbf{0\%} & 6\% & 11\% \\
\midrule
4R & 0\% & 0\% & 0\% & 0\% \\
\bottomrule
\end{tabular}
\end{table}

\subsection{Three-Way Comparison: $x{=}0$ vs.\ $x{=}1$ vs.\ PPD}
\label{app:3way}

\begin{figure*}[!ht]
    \centering
    \includegraphics[width=\textwidth]{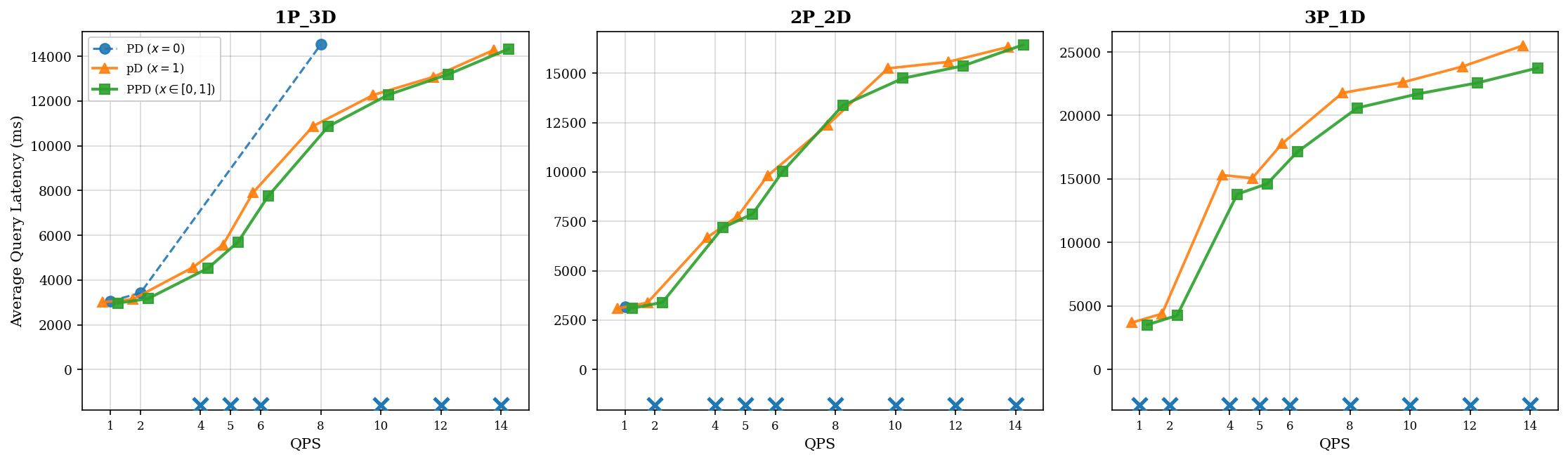}
    \caption{
        Three-way E2E latency comparison on WildChat (3 configs $\times$ 9 QPS).
        PD ($x{=}0$, blue dashed): fails catastrophically ($\times$ markers, SR$<$95\%) in 13/27 test points.
        pD ($x{=}1$, orange triangle) and PPD (green square): both maintain stability with near-identical E2E curves.
        The E2E similarity between $x{=}1$ and PPD is expected under balanced weights ($w_{\text{ttft}}{=}w_{\text{tpot}}{=}1.0$): PPD trades some TTFT for TPOT improvement on routed-back requests, which cancels out in the composite E2E metric.
        The per-metric decomposition in \Cref{tab:metric-wins} reveals that PPD breaks both static baselines' per-metric monopolies.
    }
    \label{fig:3way-e2e}
\end{figure*}

\Cref{fig:3way-e2e} presents the complete three-way comparison on WildChat across all configurations and QPS levels; the per-metric decomposition is reported as \Cref{tab:metric-wins} in the main text (\Cref{subsec:x1-comparison}).

Reading the panels left to right (1P\_3D, 2P\_2D, 3P\_1D), several patterns are visible. The static $x{=}0$ baseline (blue dashed) accumulates failure markers at progressively lower QPS as the P:D ratio shifts away from the well-balanced 1P\_3D, reflecting the diminishing decode capacity to absorb continuous KV transfers; in 3P\_1D, the single decode GPU becomes the bottleneck and a substantial fraction of the QPS range falls below the SR$\geq$95\% threshold. By contrast, both $x{=}1$ and PPD achieve a $100\%$ success rate across all 27 test points and all three configurations, and their E2E latency curves are visually almost indistinguishable across the entire QPS sweep.

This visual similarity is precisely what motivates the per-metric decomposition in the main text. Under balanced weights $\mathbf{w}=(1,1)$, PPD trades a small amount of TTFT for TPOT improvement on the subset of requests it routes back through P, and these two effects nearly cancel in the composite E2E metric, producing curves that overlay $x{=}1$'s. The per-metric numbers in \Cref{tab:metric-wins} disentangle this: PPD beats $x{=}0$ on TPOT (12 vs.\ 10 wins out of 27) and beats $x{=}1$ on TTFT (14 vs.\ 13 wins), while matching $x{=}1$'s perfect success rate. PPD does not pay a stability cost for its dynamism (\Cref{fig:3way-e2e}); the per-metric numbers in \Cref{tab:metric-wins} confirm that dynamism buys improvements no single static $x$ delivers simultaneously.

A practical reading of the three panels is that the relative value of dynamic routing scales with how stressed the decode pool is. In 1P\_3D, the abundant decode capacity makes $x{=}1$ already sufficient and PPD's per-request decisions only marginally adjust the trajectory. In 2P\_2D, the two curves remain close because the decode load is moderate. In 3P\_1D, where the lone decode node becomes the contention point under high QPS, PPD has the most room to redistribute Turn 2+ load back to the abundant P pool, narrowing the gap between TTFT-side and TPOT-side metrics. This pattern is consistent with the score-based decision rule in \Cref{eq:score}: route locally when local processing is cheap, and fall back to P when it is not.

The 30\,s timeout threshold underlying the failure markers is a deliberately conservative SLO. Tightening it to 10\,s shrinks the SR$=100\%$ region for $x{=}0$ further but leaves $x{=}1$ and PPD unchanged across the entire QPS range, because the local-cache path keeps Turn 2+ end-to-end latency well below either threshold even at QPS=14. We chose the looser threshold to make the failure mode visible across the full sweep rather than collapsing every high-QPS $x{=}0$ point into an opaque ``failed'' marker; readers prioritizing tighter SLOs can read \Cref{fig:3way-e2e} as an upper bound on $x{=}0$'s viable operating envelope.

It is also worth relating \Cref{fig:3way-e2e} to the bandwidth-simulation results in \Cref{subsec:scaling-sim}. \Cref{fig:scaling-sim} fixes QPS at 1 and varies the simulated network bandwidth, isolating the per-transfer cost component of PPD's advantage; \Cref{fig:3way-e2e} fixes bandwidth at intra-node NVLink and sweeps QPS, isolating the queuing-time component. Together the two experiments establish that PPD's gain over $x{=}0$ has two distinct origins: a static, per-request transfer-time saving that grows with slower interconnects, and a dynamic, queuing-time saving that grows with concurrent load. Neither is recoverable for $x{=}0$ once the deployment is in place, which is why even the multiplicative composition of small per-mechanism gains shows up so prominently in the aggregate E2E curves.

Finally, the 27 test points in this experiment are not chosen adversarially: each lies within the SLO envelope a reasonable operator might target, with QPS values bracketing typical conversational traffic on a 4-GPU node and conversation lengths drawn directly from real WildChat sessions. The systematic shift from $x{=}0$ failures (13/27) toward universal $x{=}1$ and PPD success (27/27) is therefore not a worst-case demonstration but a representative one. Combined with the per-metric decomposition in \Cref{tab:metric-wins}, it is the strongest single piece of evidence we present that dynamic routing is preferable to any static $x$ in production multi-turn workloads.

\end{document}